\documentclass[english,11pt,a4paper]{article}
\usepackage[margin=1in]{geometry}

\usepackage[T1]{fontenc}
\usepackage{array}
\usepackage{multirow}
\usepackage{graphicx}
\usepackage{rotating}
\usepackage{booktabs}
\usepackage{csvsimple}
\usepackage{hyperref}

\begingroup
\newcolumntype{R}[1]{>{\raggedleft\arraybackslash}p{#1}}
\newlength{\mycolwidth}
\newlength{\mysmallcolwidth}

\makeatletter

\providecommand{\\}{\\}
\newcommand{\lyxdot}{.}

\usepackage[algo2e,ruled,linesnumbered]{algorithm2e}

\@ifundefined{showcaptionsetup}{}{%
 \PassOptionsToPackage{caption=false}{subfig}}
\usepackage{subfig}
\makeatother

\usepackage{babel}
\begin{document}
\global\long\def\dist{\mathrm{dist}}

\pagenumbering{gobble}

\title{Graph Bisection with Pareto-Optimization%
\thanks{Partial support by DFG grant WA654/19-1, WA654/22-1, and Google Focused Research Award.%
}}

\author{Michael Hamann, Ben Strasser\\ 
~ \\
Institute of Theoretical Informatics,\\
Karlsruhe Institute of Technology, \\
P.O. Box 6980, 76128 Karlsruhe, Germany. \\ 
}
\date{~}

\maketitle
\begin{abstract}
We introduce FlowCutter, a novel algorithm to compute a set of edge cuts or node separators that optimize cut size and balance in the Pareto-sense. 
Our core algorithm heuristically solves the balanced connected $st$-edge-cut problem, where two given nodes $s$ and $t$ must be separated by removing edges to obtain two connected parts. 
Using the core algorithm as subroutine, we build variants that compute node separators which are independent of~$s$ and~$t$. 
From the computed Pareto-set, we can identify cuts with a particularly good trade-off between cut size and balance that can be used to compute contraction and minimum fill-in orders, which can be used in Customizable Contraction Hierarchies (CCH), a speed-up technique for shortest path computations.
Our core algorithm runs in $O(c|E|)$ time where $E$ is the set of edges and $c$ is the size of the largest outputted cut.
This makes it well-suited for separating large graphs with small cuts, such as road graphs, which is the primary application motivating our research.
For road graphs, we present an extensive experimental study demonstrating that FlowCutter outperforms the current state-of-the-art both in terms of cut sizes and CCH performance.
By evaluating FlowCutter on a standard graph partitioning benchmark, we further show that FlowCutter also finds small, balanced cuts on non-road graphs.
Another application is the computation of small tree-decompositions.
To evaluate the quality of our algorithm in this context, we entered the PACE 2016 challenge~\cite{dhjkkr-tfpac-16} and won the first place in the corresponding sequential competition track.
We can therefore conlude that our FlowCutter algorithm finds small, balanced cuts on a wide variety of graphs.
\end{abstract}

\pagenumbering{arabic}

\maketitle
\section{Introduction}

A graph cut is a set of edges, whose removal separates a graph into two sides.
Similarly, a node separator is a set of nodes whose removal separates a graph into two sides.
A cut or separator is balanced if the number of nodes in both sides is roughly the same.
Balanced graph bisection is the problem of finding a balanced cut or separator.
This is a fundamental and NP-hard~\cite{gjs-sscgp-76} graph problem that has received a lot of attention~\cite{kk-mspig-99,dgrw-gpnc-11,ss-tlagh-13,bmsw-gpgcd-13,ss-obsrn-15} and has many applications.
We present FlowCutter, a novel algorithm to compute edge cuts and node separators.
It computes a set of cuts or separators that trade-off cut respectively separator size for imbalance in the Pareto-sense.
For edge cuts, FlowCutter guarantees that the two sides of the cut are connected subgraphs.
Experimentally, we show that the computed cuts and separators have a good quality.
However, as the problem is NP-hard, we cannot prove that the cuts and separators found are always small.
The application motivating our research is accelerating shortest path computations in roads graphs~\cite{sww-daola-00,hsw-emlog-08,dgpw-crprn-13,bdgmpsww-rptn-14,dsw-cch-15}.
The main part of our paper sticks with this application and we demonstrate the usefulness of our proposed graph bisection algorithm directly by applying it in the context of shortest path computations.
However, our proposed algorithm is not specific to road graphs.
We present experiments on graphs from other applications to demonstrate this. 
Earlier versions of our work have been uploaded to ArXiv~\cite{hs-gbpo-15} and presented at the ALENEX 2016 conference~\cite{hs-gbpo-16}.

\paragraph{Contribution}

We introduce FlowCutter, a graph bisection algorithm that optimizes cut size and imbalance in the Pareto sense. 
The core FlowCutter algorithm aims to solve the balanced edge-$st$-cut graph bisection problem with connected sides. 
Using this core as subroutine we design algorithms to solve the node separator and non-$st$ variants. 
Using these we design a nested dissection-based algorithm to compute contraction node orders as needed by Customizable Contraction Hierarchies (CCH) \cite{dsw-cch-15}.
We prove that the core algorithm's running time is in $O(c|E|)$ where $E$ is the set of edges and $c$ the size of the largest cut found.
We show in an extensive experimental evaluation that FlowCutter is a good fit for road graphs, as road graphs tend to be large in terms of edge count but small in terms of cut size.
We further show that FlowCutter also finds small, balanced cuts on non-road graphs by comparing the cuts found by FlowCutter to the state-of-the-art cuts on the benchmark set maintained by Chris Walshaw~\cite{swc-acesm-04}.
Finally, we demonstrate the performance of our algorithm on a broad class of graphs, by entering and winning the corresponding sequential competition track of PACE 2016 tree-decomposition challenge.

\paragraph{Outline}

Section~\ref{sec:applications} presents an overview of related work and the core ideas of the shortest path application driving our research as well as some other applications including tree-decompositons.
Section~\ref{sec:preliminaries} presents our notation and an overview of flows, CCH, tree-decompositions, separators, and cuts and how some of these concepts relate to another. 
Section~\ref{sec:core} introduces the core FlowCutter $st$-bisection algorithm. 
Section~\ref{sec:Extensions} extends the core algorithm to general bisection, node bisection, and describes how to compute CCH contraction orders. 
Section~\ref{sec:experiments} presents an extensive experimental evaluation on road graphs against the current state of the art and on the well-known Walshaw benchmark for graph partitioning.

\section{Applications and Related Work}
\label{sec:applications}

We start by giving an overview of the core ideas employed to accelerate shortest path computations, illustrating how to apply the main topic of this work, which is graph bisection, to this application.
We limit this overview to the works immediately relevant to graph bisection and refer readers interested in further details to a recent survey article~\cite{bdgmpsww-rptn-14}.

Dijkstra's algorithm \cite{d-ntpcg-59} solves the shortest path problem in near-linear running time. 
However, this is not fast enough if the graph consists of a whole continent's road network. 
Acceleration techniques usually compute auxiliary data in a \emph{preprocessing phase} and compute the shortest paths in a \emph{query phase}.
This auxiliary data is independent of the path's endpoints and can therefore be reused for many shortest path computations. 
As roads only change slowly over time, the preprocessing phase can be slow as it does not have to be rerun very frequently.
Computing the auxiliary data usually involves optimizing some criteria, which most of the time is NP-hard \cite{bckkw-psuth-10}.
In this application, it is therefore important to produce solutions of good quality but it is not as important to do this fast.
Trading running time in the preprocessing phase for an improvement of the auxiliary data is therefore most of time worthwhile.
Some techniques such as \cite{dgpw-crprn-13,dsw-cch-15} split the preprocessing phase and introduce a \emph{customization phase}.
In the preprocessing phase only the graph but not its weights are known. 
The weights are introduced in the customization phase.
The idea behind this setup is to be able to adapt more quickly to changes in the weights, which could for example be caused by traffic congestion.

In many shortest path acceleration techniques, the preprocessing phase involves computing balanced graph edge cuts or node separators.
The central idea can be formulated in terms of edge cuts as well as node separators.
We present the node separators variant as we will evaluate our bisection algorithms using \emph{Customizable Contraction Hierarchies} (CCH) \cite{dsw-cch-15}, an accelerating technique based on the separator variant.
However, many acceleration techniques, such as \cite{dgpw-crprn-13}, are based upon the edge cut variant.
The idea can be described as follows:
Given a graph $G$ and a node separator $S$, the algorithms precompute for every node in the graph how to get to every node in $S$. 
Further, they precompute the shortest paths among all nodes of $S$.
Consider a query that asks to compute a shortest path from a node $s$ to a node $t$.
Either $s$ and $t$ are on the same or on opposite sides of $S$.
If they are on opposite sides, a shortest path can be assembled by iterating over all nodes $v$ in $S$ and combining the precomputed paths from $s$ to $v$ and from $v$ to $t$ and picking the shortest path.
The running time of a distance query in this case is thus in $O(|S|)$, which is assumed to be small for road graphs.
However, if $s$ and $t$ are on the same side then a graph search is necessary using, for example, Dijkstra's algorithm. 
On the side of $s$ and $t$ the search is unrestricted. 
However, it does not cross $S$ and instead makes use of the shortest paths precomputed between the nodes of $S$.
If the sides are of the same size and $s$ and $t$ are chosen uniformly at random then there is a 50\% probability that they are on opposite sides.
Half of the queries can therefore be answered quickly.
For the other half of queries, this approach restricts the search to half of the graph.
However, as half of a continent is still large, one usually applies this idea recursively.

The effectiveness of these techniques crucially depends on the size of the separators found.
The balance is less important.
Perfect balance is not necessary to assure that the recursion has a logarithmic depth.
This application does, however, not benefit from a perfect balance.
In practice, the contrary is true: 
Requiring perfect balance results in many small, slightly imbalanced separators not being found. 
Therefore, the perfectly balanced separators found can be larger.
This larger size is detrimental to the running time of the query phase, compared to using the smaller slightly imbalanced separators.
Fortunately, road graphs have small separators and cuts because of geographical features such as rivers or mountains. 
Previous work has coined the term \emph{natural cuts} for this phenomenon \cite{dgrw-gpnc-11}. 
However, identifying these natural cuts is a difficult problem.

Graph partitioning software used for road graphs include \emph{KaHip} \cite{ss-tlagh-13}, \emph{Metis} \cite{kk-mspig-99}, \emph{InertialFlow} \cite{ss-obsrn-15}, and \emph{PUNCH} \cite{dgrw-gpnc-11}.
We experimentally compare FlowCutter with the first three. 
As we unfortunately have no implementation of PUNCH, we omitted an experimental comparison with it.
All of these works formalize the graph bisection problem as a bicriteria problem optimizing the cut size and the imbalance. 
The \emph{imbalance} measures how much the sizes of both sides differ and is small if the sides are balanced. 
The standard approach is to bound the imbalance and minimize the cut size. 
However, this approach has several shortcomings. 
Consider a graph with a million nodes and set the maximum imbalance to 1\%.
Suppose an algorithm finds a cut $C_{1}$ with 180 edges and 0.9\% imbalance. 
This is all the information of the cut's quality that is provided.
Can you decide solely based on this information, whether this is a good cut? 
It seems good as 180 is small compared to the node count.
However, we would come to a different conclusion, if we knew that a cut $C_{2}$ with 90 edges and 1.1\% imbalance existed. 
In our application --- shortest paths --- moving a few nodes to the other side of a cut is no problem. 
However, halving the cut size has a huge impact. 
The cut $C_{2}$ is thus clearly superior. 
Further, assume that a third cut $C_{3}$ with 180 edges and 0.7\% imbalance existed. 
$C_{3}$ dominates $C_{1}$ in both criteria. 
However, both are equivalent with respect to the standard problem formulation and thus a program is not required to output $C_{3}$ instead of $C_{1}$. 
To overcome these problems, our approach computes a set of cuts that optimize cut size and imbalance in the Pareto sense, i.e., it tries to compute solutions that are Pareto-optimal.
As this problem is NP-hard, one cannot expect to always find the exact Pareto curve. 
A further significant shortcoming of the state-of-the-art partitioners, with the exception of InertialFlow, is that they were designed for small imbalances. 
Common benchmarks, such as the one maintained by Chris Walshaw~\cite{swc-acesm-04}, only include test cases with imbalances up to 5\%. 
However, for our application imbalances of 50\% can be fine.
For such high imbalances unexpected things happen with the standard software, such as increasing the allowed imbalance can increase the achieved cut sizes.
Indeed, KaHip, one of the competitors, has been updated, as reaction to the conference version of our work~\cite{hs-gbpo-16}, to overcome some of these shortcomings.
The newer version produces better results for higher imbalances than the old version.

We use Customizable Contraction Hierarchies (CCH) \cite{dsw-cch-15} to evaluate the quality of the separators found by FlowCutter. 
We present a more detailed CCH-overview in Section~\ref{sec:cch}.
The CCH-auxiliary data at its core is a chordal supergraph of the road graph.
The maximum cliques of $G'$, which can be efficiently identified in chordal graphs, correspond to the bags of a tree-decomposition.
This connection gives us a bridge to the vast field of tree-decomposition theory.
In Section~\ref{sec:tree-decomp}, we give a high-level overview of the concepts related to tree-decompositions and CCH.
For a more in-depth survey we refer to~\cite{bp-aicgc-93},~\cite{b-atgt-93} and~\cite{b-tsa-07}.
In some works tree-decompositions are also called clique-trees.

\paragraph{Other Applications}

A vast amount of algorithms for NP-hard graph problems exist that are fixed-parameter tractable in the tree-width of a graph $G$~\cite{c-tir-90,b-tsa-07}.
It is therefore an interesting question whether algorithms being able to compute good tree-decompositions in practice leads to practicable variants of these algorithms.
To investigate this question, the PACE competition~\cite{dhjkkr-tfpac-16} was held in 2016 at IPEC, a conference with a focus on fixed-parameter tractable algorithms.
The objective of two competition tracks was to compute a small tree decomposition within a limited time frame.
The tracks differed in whether parallelism was allowed or not.
To evaluate the performance of FlowCutter in this context, we submitted our algorithm.
Our implementation runs FlowCutter iteratively with varying parameters until the time limit is reached. 
The parallel version runs several FlowCutter instances in parallel.
The code submitted to the PACE challenge is open source and available at~\cite{pace-code}.
In the sequential track our implementation won the first place out of six submissions and in the parallel track it won the second place out of 3 submissions.
Given these results, it is safe to say, that our algorithm is at least highly competitive, if not the state-of-the-art, in terms of computing tree decompositions in practice.

The contraction orders used by CCH, which as based upon nested dissection \cite{g-ndrfe-73,lrt-gnd-79}, are also called minimum fill-in orders in the context of sparse matrices.
This establishes a connection to the theory of quickly solving sparse systems of linear equations.
Indeed, METIS was developed with this application in mind~\cite{kk-mspig-99}.
METIS was not developed to bisect road graphs.
The fact that we use METIS in the context of road graphs is therefore an example of this theoretical connection being exploited in practice.
Using the same connection, it is also possible to use FlowCutter to solve sparse system of linear equations. 
However, even though these two applications are so closely related, the precise trade-off between the various optimization criteria differs.
For example in the context of sparse equation systems, cut size is less important than in the road setting whereas having a small bisection algorithm running time is more important.

Another application is information propagation in belief networks~\cite{hd-ijarv-96}. 
In this setting, a set of random variables is given. 
It is known how these random variables depend on each other and their interactions are modeled as a graph whose nodes are the random variables.
The question is how the distributions change throughout the graph if the distribution of a subset of the variables changes, i.e., some but not all variables are measured.
To solve this problem, so called junction-trees are employed.
Junction-trees are essentially another name for tree-decompositions.
As we can use FlowCutter to compute tree-decompositions, we can also use it to compute junction-trees.

\section{Preliminaries}
\label{sec:preliminaries}

A \emph{directed graph} is denoted by $G=(V,A)$ with \emph{node set} $V$ and \emph{arc set} $A\subseteq V\times V$. 
Similarly, an \emph{undirected} graph is denoted by $G=(V,E)$ with node set $V$ and \emph{edge set} $E\subseteq \{e \in 2^V : |e| = 2\}$.
Arcs have an implicit direction, whereas edges are undirected.
A directed graph is \emph{symmetric}, if for every arc $(y,x)$ there exists an arc $(x,y)$.
In a slight abuse of notation, we do not discern between undirected and directed, symmetric graphs.
We identify an edge $\{x,y\}$ with the corresponding pair of arcs $(x,y)$ and $(y,x)$.
We set $n:=\left|V\right|$ and $m:=\left|A\right|$.
As input, we only consider undirected graphs without multi-edges and without reflexive loops, i.e., without arcs of the form $(x,x)$.
Road graphs that do not fit this description are modified\footnote{As shown in the CCH-paper~\cite{dsw-cch-15}, this simplification does not hinder the applicability of CCH to directed graphs.} by removing multi-edges, removing reflexive loops, and adding backarcs in the case of one-way streets.
In intermediate steps of our algorithm, we also consider non-symmetric directed graphs.
The \emph{out-degree} $d_{o}(x)$ of a node $x$ is the number of outgoing arcs. 
Similarly, the \emph{in-degree} $d_{i}(x)$ is the number of incoming arcs. 
In symmetric graphs, we refer to the value as \emph{degree} $d(x)$ of $x$, as $d_{i}(x)=d_{o}(x)$. 
A \emph{degree-2-chain} is a sequence of adjacent nodes $x,y_{1}\ldots y_{k},z$ in a symmetric graph such that $k\ge1$, $d(x)\neq 2$, $d(z)\neq 2$, and $\forall i:d(y_{i})=2$.
An \emph{$xy$-path} $P$ is a list $(x,p_{1}),(p_{1},p_{2})\ldots(p_{i},y)$ of adjacent arcs and $i+1$ is $P$'s length. 
The \emph{distance} $\dist(x,y)$ is defined as the minimum length over all $xy$-paths. 

\subsection{Cuts and Separators}

A \emph{cut} $(V_{1},V_{2})$ is a partition of $V$ into two disjoint sets $V_{1}$ and $V_{2}$ such that $V=V_{1}\cup V_{2}$. 
An arc $(x,y)$ with $x\in V_{1}$ and $y\in V_{2}$ is called \emph{cut-arc}.
In another slight abuse of notation, we do not discern between the node partition and the set of cut-arcs.
The \emph{size of a cut} is the number of cut-arcs. 
A min-cut is a cut of minimum size.
A \emph{separator} $(V_{1},V_{2},Q)$ is a partition of $V$ into three disjoint sets $V_{1}$, $V_{2}$ and $Q$ such that $V=V_{1}\cup V_{2}\cup Q$. 
There must be no arc between $V_{1}$ and $V_{2}$. 
The cardinality of $Q$ is the \emph{separator's size}.
The \emph{imbalance} $\epsilon$ of a cut or separator is defined as the smallest number such that $\max\left\{ \left|V_{1}\right|,\left|V_{2}\right|\right\} \le\left\lceil (1+\epsilon)n/2\right\rceil $.
The imbalance of a separator is defined analogously. 
For edge cuts $0\le \epsilon \le 1$ holds. 
This is not necessarily the case for node separators.
The separator itself may contain nodes, making it possible that the minimum $\epsilon$ is smaller than $0$, as both sides can have fewer than $n/2$ nodes.
An \emph{$ST$-cut/separator} is a cut/separator between two disjoint node sets $S$ and $T$ such that $S\subseteq V_{1}$ and $T\subseteq V_{2}$.
If $S=\{s\}$ and $T=\{t\}$, we write \emph{$st$-cut/separator}.
The \emph{expansion} of a cut/separator is the cut's size divided by $\min\{\left|V_{1}\right|,\left|V_{2}\right|\}$. 

\paragraph{Pareto-Optimization and NP-hardness}

Computing cuts (and separators) is inherently a bicriteria problem: We want to minimize the cut size and minimize the imbalance. 
A cut $C_1$ dominates a cut $C_2$ if $C_1$ is strictly better with respect to one criterion and no worse with respect to the other criterion. 
A cut that is not dominated by any other cut is \emph{Pareto-optimal}. 
We refer to the pair of imbalance and cut size of a Pareto-optimal cut as \emph{Pareto-trade-off}. 
It is possible that several Pareto-optimal cuts exist with the same trade-off. 
The problem we consider asks to compute one cut for every Pareto-trade-off.
If there are several, then the algorithm is free to pick any one of them.

This is a departure from existing experimental papers~\cite{kk-mspig-99,swc-acesm-04,ss-tlagh-13,bmsw-gpgcd-13,dfgrw-a-14,w-fsns-14,ss-obsrn-15} that consider the problem of finding a smallest cut subject to an imbalance bounded by an input parameter. 
Given a cut for every Pareto-trade-off, it is easy to find a smallest cut with a bounded imbalance. 
However, a cut with minimum size with an imbalance bounded by an input parameter is not necessarily Pareto-optimal: It is possible that a more balanced cut with the same size exists. 
Our problem setting is therefore a strict generalization of the problem setting considered in previous works.

The minimum cut problem disregarding the imbalance is solvable in polynomial time~\cite{ff-mftn-56}. 
However, nearly all cut-problems that combine optimizing imbalance and cut size are NP-hard. 
Examples include:
\begin{itemize}
\item Finding a perfectly balanced minimum cut, i.e., one with $\epsilon=0$, is NP-hard~\cite{gjs-sscgp-76}. 
\item A \emph{sparsest cut} $C$ is a cut that minimizes $\frac{|C|}{|V_1|\cdot |V_2|}$.
A sparsest cut is Pareto-optimal.
Finding a sparsest cut is NP-hard~\cite{lr-m-99}. 
\item Even computing, for a fixed $st$-pair, a most balanced cut among all $st$-cuts of minimum size is already NP-hard~\cite{b-m-10}.
\item In \cite{ww-bmcgb-93}, it was shown that computing a minimum cut that respects a given imbalance is NP-hard.
\end{itemize}
Being able to compute a cut for every Pareto-trade-off efficiently would yield an efficient algorithm for all these NP-hard problems.
Unless P=NP, we can therefore not hope to find an efficient algorithm that provably computes an optimal cut for every Pareto-trade-off. 
Our algorithm tries to heuristically compute in a single run a cut for every Pareto-trade-off.

\subsection{Flows}

In this paper we only consider \emph{unit flows}.
These are a restricted variant of the flow problem: Every arc has capacity 1 and an integral flow intensity of either 0 or 1. 
Formally, a flow is a function $f:A\rightarrow\{0,1\}$. 
An arc $a$ with $f(a)=1$ is \emph{saturated}. 
Denote by $p(x)=\sum_{(x,y)\in A}f(x,y) - \sum_{(y,x)\in A}f(y,x)$ the \emph{surplus of a node} $x$. 
A flow is valid with respect to a source set $S$ and target set $T$ if and only if: 
\begin{itemize}
\item Flow may be created at sources, i.e., $\forall s\in S:p(s)\ge0$, 
\item flow may be removed at targets, i.e., $\forall t\in T:p(t)\le0$, 
\item flow is conserved at all other nodes, i.e., $\forall x\in V\backslash(S\cup T):p(x)=0$,
\item and flow does not flow in both directions, i.e., for all $(x,y)\in A$ such that $(y,x)\in A$ exists, it holds that $f(x,y)=0\vee f(y,x)=0$.
\end{itemize}
The flow \emph{intensity} is defined as the sum over all $f(x,y)$ for arcs $(x,y)$ with $x\in S$ and $y\not\in S$. 
In other works, the flow intensity is sometimes also called flow value.
A path $a_{1},a_{2}\ldots,a_{i}$ is \emph{saturated} if there exists an $i$ with $f(a_{i})=1$.
A node $x$ is \emph{source-reachable} if a non-saturated $sx$-path exists with $s\in S$. 
Similarly, a node $x$ is called \emph{target-reachable} if a non-saturated $xt$-path exists with $t\in T$.
We denote by $S_{R}$ the set of all \emph{source-reachable nodes} and by $T_{R}$ the set of all \emph{target reachable nodes}. 
In~\cite{ff-mftn-56} it was shown that a flow is maximum, if and only if no non-saturated $st$-path with $s\in S$ and $t\in T$ exists. 
If such a path exists, then it is called \emph{augmenting path}. 
The classic approach to compute max-flows consists of iteratively searching for augmenting paths. 
Our algorithm builds upon this classic approach. 
The minimum $ST$-cut size corresponds to the maximum $ST$-flow intensity. 
We define the \emph{source-side cut} as $(S_{R},V\backslash S_{R})$ and the \emph{target-side cut} as $(T_{R},V\backslash T_{R})$.  
Note that in general max-flows and min-cuts are not unique. 
However, the source-side and target-side cuts are.
The source-side and target-side cuts are the same for every max-flow.
The later is an implication of Corollary 5.3 of \cite{ff-fn-62}, which states that every min-cut is saturated with respect to every max-flow.

\subsection{Customizable Contraction Hierarchy}
\label{sec:cch}

A Customizable Contraction Hierarchy (CCH) is an acceleration algorithm for shortest path computations.
We only give a high-level overview, as we use CCH only to evaluate the quality of our separators.
No part of FlowCutter builds upon CCH.
The CCH details are in explained in~\cite{dsw-cch-15}.
The technique uses three phases, i.e., a preprocessing phase, a customization phase, and a query phase.
In the preprocessing phase, the arc weights are unknown.
These are integrated into the auxiliary data in the customization phase.
Shortest paths are computed in the query phase.

\paragraph{Preprocessing}
The name-giving operation is the node contraction: 
Contracting a node $v$ consists of removing $v$ and adding edges between all of $v$'s neighbors, if they did not already exist.
The input to CCH consists of a node \emph{contraction order} along which the nodes are iteratively contracted.
This yields a supergraph $G'$ of the input graph.
By convention, we say for every edge $\{x,y\}$ where $x$ is contracted before $y$, that $x$ is lower than $y$.
The position of a node in the order is called \emph{rank}.

\paragraph{Customization}
The weights of $G'$ are computed using an algorithm that essentially enumerates all triangles in $G'$.
Initially, all edges of $G'$ already present in $G$ are assigned their input weights, whereas all edges introduced during the contraction are giving the weight $\infty$.
The algorithm then enumerates all triangles $\{x,y,z\}$ in $G'$ ordered increasingly by the position of the lowest node in the triangle.
Suppose that $z$ is this lowest node.  
For each triangle the algorithm executes $w(x,y)\leftarrow \min \{w(x,y),w(x,z)+w(z,y)\}$.
As shown in~\cite{dsw-cch-15}, for every pair of nodes $s$ and $t$ there is a shortest $st$-\emph{up-down-path} after the customization has finished.
This is a path with nodes $v_1\ldots v_m \ldots v_\ell$ such that the nodes $v_1\ldots v_m$ appear ordered increasingly by rank and the nodes $v_m\ldots v_\ell$ appear ordered decreasingly by rank.
The nodes $v_1\ldots v_m$ form the \emph{upward part} of the path and the nodes $v_m\ldots v_\ell$ form the \emph{downward part}.

\paragraph{Query}
Given the weights of $G'$, the \emph{shortest path query} consists of a bidirectional search in $G'$.
Both the forward and the backward searches follows only upward arcs $(x,y)$ such that $x$ has a lower rank than $y$.
The forward search finds the upward part and the backward search the downward part of a shortest $st$-up-down-path.
The \emph{search space} of a node $z$ is the subgraph of $G'$ that is reachable from $z$ while only following upward arcs.
The query therefore only explores at most the whole search spaces of $s$ and $t$.

\paragraph{Performance}
Smaller search spaces yield faster queries. 
Fewer triangles in $G'$ yield a faster customization.
Fewer edges in $G'$ result in less memory consumption.
All these quality metrics depend on the contraction order.

We compute contraction orders using \emph{nested dissection}~\cite{g-ndrfe-73,lrt-gnd-79}, which works as follows: (1) Determine a small balanced separator $S$, (2) recursively compute orders $L$ and $R$ for both sides, (3) the contraction order of $G$ consists of first contracting the nodes along $L$, then $R$, and finally along an arbitrary order of the nodes in $S$.  
The quality of the so obtained contraction order depends on the quality of the separators used in its construction.
Finding these separators is where FlowCutter fits into the big picture.

\subsection{Tree-Decompositions}
\label{sec:tree-decomp}

We present an overview of the theory of tree-decompositions.
As the sheer quantity of works written on this subject makes it impossible to introduce them all, we refer the interested reader to three survey articles~\cite{bp-aicgc-93,b-atgt-93,b-tsa-07}.

\paragraph{Definition}

A \emph{tree-decomposition} of an undirected graph $G=(V,E)$ is a pair $(B,T)$ where $B$ is a set of subsets of $V$.
The elements of $B$ are called \emph{bags}.
Every bag is a set of nodes.
The union of all bags must equal $V$.
It is required that for every $\{x,y\}\in E$ there is a bag $b\in B$ such that $x\in b$ and $y\in b$, i.e., every edge is in one or more bags.
$T$ is a tree with the elements of $B$ as nodes. 
$T$ is called the \emph{backbone} of the decomposition.
Finally, it is required that for every pair of bags $b_s$ and $b_t$ whose intersection contains a vertex $v \in b_s \cap b_t$, that all bags $b_i$ on the unique path from $b_s$ to $b_t$ in the backbone $T$ also contain $v$.
The \emph{width} of a decomposition is defined as the maximum bag size minus one.
The \emph{tree-width} of a graph is the minimum width over all decompositions.
For simplicity, we will further assume in our overview that all tree-decompositions are well behaved, i.e., the subgraph of $G$ induced by a bag is connected and no bag is a subset of another bag.
Given a tree-decomposition that does not have these properties it is possible to obtain one that does by removing superfluous bags and splitting disconnected bags. 

\paragraph{Chordal Graphs and Perfect Elimination Orders}
An undirected graph $G$ is \emph{chordal}, if in every cycle $C$ of at least four nodes there is a chord, i.e., a pair of nodes that are adjacent in the graph $G$ but not adjacent in the cycle $C$. 
A node is call \emph{simplicial} if its neighbors form a clique.
Every chordal graph contains at least one simplicial node \cite{fg-imig-65}. 
A perfect elimination order of an undirected graph $G$ is a node order $v_1,v_2\ldots v_n$, such that $v_i$ is simplicial in the subgraph of $G$ induced by $v_i,v_{i+1}\ldots v_n$. 
Not every graph possesses a perfect elimination order.
Chordal graphs can be characterized as the graphs that possess a perfect elimination order \cite{fg-imig-65}.
Such an order can be constructed, given a chordal graph, by iteratively removing simplicial nodes from a chordal graph.
As chordal graphs can have several simplicial nodes, chordal graphs can possess several perfect elimination orders.

The CCH contraction order is a perfect elimination order of the supergraph $G'$ constructed in the CCH.
$G'$ is thus a chordal supergraph of the input graph $G$.
In the context of chordal graphs, a contraction order is also called \emph{elimination order}.

\paragraph{Converting Structures}

A tree-decomposition can be constructed, given a chordal supergraph $G'$ of a graph $G$ as follows:
Start by identifying the maximal cliques in $G'$.
This can be done in polynomial running time using the perfect elimination ordering $v_i$. 
Every maximal clique of $G'$ must appear as union of a node $v_i$ with its neighborhood in the subgraph of $G'$ induced by $v_i,v_{i+1}\ldots v_n$.
Testing which of these linearly many neighborhoods are maximal cliques results in all maximal cliques being found.
The maximal cliques of $G'$ are the bags of the decomposition.
We denote their set by $B$.
These cliques are the nodes of the tree-backbone $T$.
It remains to construct the edges of the tree-backbone $T$.
We therefore consider the weighted undirected graph $T'=(B,E)$, where an edge $\{x,y\}$ exists if the intersection of bags $x$ and $y$ is non-empty.
An edge $\{x,y\}$ is weighted by the number of nodes in the intersection of $x$ and $y$.
$T'$ is not necessarily a tree and thus not a valid backbone.
However, every maximum spanning tree of $T'$ is a valid tree-backbone~\cite{bg-pns-81}.
Computing a maximum spanning tree thus completes the tree-decomposition construction.
Given a tree-decomposition we can easily get back to the corresponding chordal graph.
The transformation consists of adding edges between all nodes in each bag.

This gives us three views that encode essentially the same information: Chordal supergraphs, tree-decompositons, and elimination orders.
These are interconvertible as follows:
\begin{itemize}
\item Given an undirected graph $G$ and an elimination order $o$ we can get to a chordal supergraph $G'$ by iteratively contracting the nodes.
\item By computing a perfect elimination order of $G'$ we can obtain an elimination order $o'$ of $G$. The orders $o$ and $o'$ are not necessarily the same, but they induce the same chordal supergraph.
\item From the chordal supergraph $G'$ we can construct a tree-decomposition of $G$ as described above.
\item From the tree-decomposition of $G$ we can get back to the chordal supergraph $G'$ by adding edges in each bag.
\end{itemize}
It is guaranteed that none of these transformations increases the width of the corresponding tree-decomposition.

\paragraph{Elimination Tree}

With respect to an elimination order we can define the \emph{elimination tree}\footnote{The elimination tree is actually a forest, if the input graph is disconnected.}.
For every node $v_i$ at position $i$ in the order, we define its parent in the elimination tree as the first node $v_j$ that appears in the order after $v_i$ such that an edge $\{v_i,v_j\}$ exists.
If no such $v_j$ exists, the node is a root in the elimination tree.

It can be shown that the nodes in the search space of $v$ is the set of ancestors of $v$ in the elimination tree~\cite{bcrw-sssch-13}.
In a shortest path distance query from $s$ to $t$ in a CCH, the ancestors of $s$ and $t$ in the elimination tree are therefore the set of nodes touched by the CCH query algorithm.
It is therefore our goal to minimize the depth of the elimination tree.

Two elimination orders that yield the same chordal supergraph do not necessarily yield the same elimination tree.
The minimum depth over the elimination trees over all elimination orders is called the \emph{tree-depth}. 

\paragraph{Computing Elimination Orders}

Commonly used algorithms to compute tree-decompositions of large graphs in practice are heuristics that try to guess the elimination order.
The simplest is the minimum degree heuristic \cite{m-teial-57,gl-temdo-89}.
It consists of iteratively contracting a node of minimum degree.
This simple algorithm is usually fast and works well enough on some graphs but at least on road graphs there is a large gap to the optimal orders~\cite{dsw-cch-15}.

A more sophisticated approach is called \emph{nested dissection}~\cite{g-ndrfe-73,lrt-gnd-79}.
The idea is precisely the same as the approach we used to compute the CCH contraction orders.
It consists of finding a small balanced separator and placing these nodes at the end of the elimination order.
The separator is then removed from the graph and the algorithm recursively continues on both sides.

The operation of contracting a separator last can be interpreted as trying to guess a central edge in the tree-backbone.
Denote by $a$ and $b$ two non-leaf bags and by $\{a,b\}$ an edge in the backbone.
The nodes of $a\cap b$ form a separator of the input graph \cite{bp-aicgc-93}.
If the edge $\{a,b\}$ is positioned near the center of the tree-backbone then the sides of the separator $a\cap b$ are likely of roughly the same size.
This motivates why nested dissection works in practice.

\subsection{Bounding CCH Performance in Terms of Tree-Width and -Depth}

We can express the CCH performance in terms of tree-depth, which can be bounded using the tree-width.

A tree-decomposition of minimum width does not necessarily have an elimination order that results in a minimum elimination tree depth.
For example, a path has a tree-decomposition of width 1 as it is a tree.
However, no elimination order exists that yields a depth smaller than $n/2$.
Fortunately, a tree-decomposition exists with a width in $O(\log n)$ and a depth in $O(\log n)$.
It can be obtained using nested dissection with balanced separators.  
Besides small bags, a tree-decomposition must also have a logarithmic diameter to allow for a low elimination tree depth.
This logarithmic diameter corresponds to the logarithmic recursion depth obtained by recursively bisecting a graph along a balanced separator.

Denote by $\mathrm{tw}$ the tree-width of the input graph $G$ and by $\mathrm{td}$ its tree-depth.
\cite{bghk-atpfs-95} have shown that there always exists an elimination order of $G$ that yields an elimination tree depth of $O(\mathrm{tw}\log n)$ but the corresponding tree-decomposition does not necessarily have a minimum width.
Fortunately, the depth of every elimination tree is an upper bound to the width of the corresponding tree-decomposition.
There is therefore always a tree-decomposition with width $O(\mathrm{tw}\log n)$ that admits an elimination tree of depth $O(\mathrm{tw}\log n)$.
Assuming that we could construct an elimination order $\Pi$ of minimum elimination tree depth, which is NP-hard \cite{p-t-88}, we could bound the performance of the corresponding CCH in terms of $\mathrm{td}$.

A CCH-query from $s$ to $t$ explores the search spaces of $s$ and $t$.
The number of nodes in each of them is bounded by $\mathrm{td}$ and thus no more than $2\mathrm{td}$ nodes are visited.
The running time of this exploration is however only bounded by $O(\mathrm{td}^2)$ as the subgraphs can be dense.
In practice the average depth and the average number of arcs over all nodes might be a better performance indicator than the maximum numbers.
Fortunately, using a simple algorithm that proceeds top-down along the elimination tree we can compute the average and maximum number of nodes and arcs in the subgraph of each node efficiently.

We are further interested in the number of edges and triangles in $G'$ as these correspond to the customization running times and the memory consumption, respectively.
The number of edges is equal to $\sum_{v} d_o(v)$ when directing every edge of $G'$ upward.
The number of triangles in which a node $v$ appears as lowest node is $(d_o(v)\cdot (d_o(v)-1))/2$ as the upper neighbors of $v$ form a clique.
The total number of triangles is therefore $\sum_{v} (d_o(v)\cdot (d_o(v)-1))/2$.

We can bound $d_o(v)$ using the width of the tree-decomposition corresponding to $\Pi$. 
Recall that this decomposition does not necessarily have a minimum width, but, as shown by \cite{bghk-atpfs-95}, its width can be bounded by $\mathrm{td}$.
We thus obtain the bounds of $O(n \mathrm{td})$ for the number of edges and $O(n \mathrm{td}^2)$ for the number of triangles.
Fortunately, for planar graphs \cite{lrt-gnd-79,gt-t-86} have shown that, there exist nested dissection orders such that, the number of edges in $G'$ is bounded by $O(n \log n)$.
This is usually smaller than $O(n \mathrm{td})$ upper edge count bound.
Further, this formula can also be used to bound the number of triangle by $O(n \mathrm{td} \log n)$ as follows: We can bound $\sum_v d_o^2(v)$ by $(\max_v d_o(v))\cdot\sum_v d_o(v)$. $\max_v d_o(v)$ is at most $\mathrm{td}$ and $\sum_v d_o(v)$ is the number of edges, i.e., at most $O(n\log n)$.
Unfortunately, \cite{lrt-gnd-79,gt-t-86} do not analyze the corresponding elimination tree heights.
It is thus unknown whether they are in $O(\mathrm{td})$.
Road graphs are not strictly planar, but often planar enough to make this result relevant.

\section{Core FlowCutter Algorithm}
\label{sec:core}

\begin{figure}
\begin{center}
\subfloat[\label{fig:balanced-cut}Balanced cut $C$]{\begin{centering}
\includegraphics[width=0.3\textwidth,page=1]{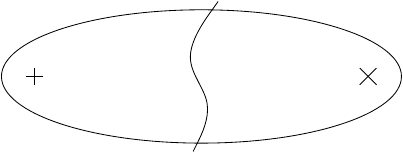}
\end{centering}

}~\subfloat[\label{fig:unbalanced-cut}Unbalanced cut $C$]{\centering{}\includegraphics[width=0.3\textwidth,page=2]{graph}}~\subfloat[\label{fig:extra-Sources}Extra sources to avoid $C$]{\begin{centering}
\includegraphics[width=0.3\textwidth,page=3]{graph}
\end{centering}

}
\end{center}

\begin{center}
\subfloat[\label{fig:next-source-side}Source side cut $C'$]{\begin{centering}
\includegraphics[width=0.3\textwidth,page=4]{graph}
\end{centering}

}~\subfloat[\label{fig:next-target-side}Target side cut $C'$]{\begin{centering}
\includegraphics[width=0.3\textwidth,page=5]{graph}
\end{centering}

}
\end{center}

\caption{\label{fig:cut-example}An ellipse represents a graph and the curved
lines are cuts. The ``+''-signs represent source nodes and ``$\times$''-signs
represent target nodes.}
\end{figure}

In the previous two sections, we described how finding good graphs cuts and separators is beneficial to many applications.
In this section, we propose our novel algorithm to compute graph cuts named FlowCutter.

FlowCutter works by computing a sequence of $st$-cuts of increasing size.
The more imbalanced cuts are computed first and are followed by more balanced ones.
The cuts in this sequence form, after removing dominated ones, the heuristically approached Pareto-set.
During its execution our algorithm maintains a maximum flow.
With respect to this flow there is a source-side cut $C_S$  and a target-side cut $C_T$.
Our algorithm picks one of the two as the next cut $C$ that it inserts into the set. 
After choosing $C$ it modifies the set of source and target nodes and potentially augments the maintained flow.
This results in a new pair of source-side and target-side cuts.
FlowCutter picks $C_S$ as $C$ if there are less or equally many nodes on the source side of $C_S$ than there are on the target side of $C_T$.

Consider the situation depicted in Figure~\ref{fig:cut-example}.
Initially $s$ is the only source node and $t$ is the only target node.
Our algorithm starts by computing a maximum $st$-flow.
If we are lucky and the cut $C$ is perfectly balanced as in Figure~\ref{fig:balanced-cut} then our algorithm is finished. 
However, most of the time we are unlucky and we either have the situation depicted in Figure~\ref{fig:unbalanced-cut} where the source's side of $C$ is too small or the analogous situation where the target's side of $C$ is too small. 
Assume without loss of generality that the source's side is too small. 
Our algorithm now transforms non-source nodes into additional source nodes to invalidate $C$ and computes a new more balanced $st$-min-cut $C'$, the second cut in the sequence. 
To invalidate $C$, our algorithm does two things:
It marks all nodes on the source's side of $C$ as source nodes and marks one node as source node on the target's side of $C$ that is incident to a cut edge. 
This node on the target's side is called the \emph{piercing node} and the corresponding cut arc is called \emph{piercing arc}. 
The situation is illustrated in Figure~\ref{fig:extra-Sources}.
All nodes on the source's side are marked as source node to assure that $C'$ does not cut through the source's side. 
The piercing node is necessary to assure that $C'\neq C$. 
Choosing a good piercing arc is crucial for good quality. 
In this section, we assume that we have a \emph{piercing oracle} that determines the piercing arc given $C$ in time linear in the size of $C$. 
In Section~\ref{sec:Pierce-Heuristic} we describe heuristics to implement such a piercing oracle. 
For the algorithm to make progress we need that $C'$ is non-dominated.
As its size is at least the size of $C$, this is equivalent with $C'$ being more balanced than $C$.
However, we can only guarantee this if $C'$ is, just as $C$, a source-side cut as in Figure~\ref{fig:next-source-side}.
If $C'$ is a target-side cut as in Figure~\ref{fig:next-target-side} then $C'$ might have a worse balance than $C$.
Luckily, as our algorithm progresses, either the target side will catch up with the balance of the source side or another source side cut is found.
In both cases our algorithm eventually finds a cut with a better balance than~$C$. 

Our algorithms grows the sides around the source and target nodes.
By doing so it can guarantee that both sides are connected.
In some applications, this is a desired property.
In others, it might be an obstacle to finding the smallest possible cuts.
Depending on the application this property is therefore either a feature or a drawback of our algorithm.

\begin{algorithm2e}[t]

$S\leftarrow\{s\}$; $T\leftarrow\{t\}$\;
$S_R\leftarrow S$; $T_R\leftarrow T$\; 
forward-grow $S_R$; backward-grow $T_R$\;

\While{$S\cap T = \emptyset$}{
  \uIf{$S_{R}\cap T_{R}\neq\emptyset$}{
    augment flow by one\;
    $S_R\leftarrow S$; $T_R\leftarrow T$\; 
    forward-grow $S_R$; backward-grow $T_R$\;

  }
  \Else{
    \uIf{$|S_{R}|\le |T_{R}|$}{
      forward-grow $S$\;
      \tcp{now $S=S_R$}
      output source side cut edges\;
      $x\leftarrow$ pierce node\;
      $S\leftarrow S\cup \{x\}$;
      $S_R\leftarrow S_R\cup \{x\}$\;
      forward-grow $S_R$\;
    }
    \Else{
      \tcp{Analogous for target side}
    }
  }
}
\caption{Pseudo-Code illustrating the core $st$-bisection algorithm.}
\label{fig:code}
\end{algorithm2e}

Our algorithm computes the $st$-min-cuts by finding max-flows and using the max-flow-min-cut duality~\cite{ff-mftn-56}. 
It assigns unit capacities to every edge and computes the flow by successively searching for augmenting paths. 
A core observation of our algorithm is that turning nodes into sources or targets never invalidates the flow. 
It is only possible that new augmenting paths are created increasing the maximum flow intensity. 
Given a set of nodes $X$ we say that \emph{forward growing} $X$ consists of adding all nodes $y$ to $X$ for which a node $x\in X$ and a non-saturated $xy$-path exist. 
Analogously, \emph{backward growing} $X$ consists of adding all nodes $y$ for which a non-saturated $yx$-path exists. 
The growing operations are implemented using a graph traversal algorithm (such as a DFS or BFS) that only follows non-saturated arcs. 
The algorithm maintains besides the flow values four node sets: the set of sources $S$, the set of targets $T$, the set source-reachable nodes $S_{R}$, and the set of target-reachable nodes $T_{R}$.
An augmenting path exists if and only if $S_{R}\cap T_{R}\neq\emptyset$.
Initially, we set $S=\{s\}$ and $T=\{t\}$. 
Our algorithm works in rounds. 
In every round it tests whether an augmenting path exists.
If one exists, the flow is augmented and $S_{R}$ and $T_{R}$ are recomputed. 
If no augmenting path exists, then it must enlarge either $S$ or $T$. 
This operation also yields the next cut. 
It then selects a piercing arc and grows $S_{R}$ and $T_{R}$ accordingly. 
The pseudo-code is presented as Algorithm~\ref{fig:code}.

\subsection{Running Time.}

Assuming a piercing oracle with a running time linear in the current cut size, we can show that the algorithm has a running time in $O(cm)$ where $c$ is the size of the most balanced cut found and $m$ is the number of edges in the graph. 
The exact details are slightly more involved but, fortunately, the core argument is simple. 
All sets only grow unless we find an augmenting path. 
As each node can only be added once to each set, the running time between finding two augmenting paths is linear.
In total, we find $c$ augmenting paths.
The total running time is thus in $O(cm)$.
The remainder of this section contains the details necessary to formally show the $O(cm)$ worst case running time.

The lines 1-3 of Algorithm~\ref{fig:code}, which initialize the data structures, have a running time in $O(m)$ and are therefore unproblematic.
The condition in line 4 can be implemented in $O(1)$ as follows:
$S$ and $T$ only grow. 
Using two bit-arrays with $n$ elements we can store which node is in $S$ and which in $T$.
When adding a node, we raise the corresponding bit and check whether the bit in the array is set.
As $S$ and $T$ only grow, the loop will abort the next time line 4 is reached, once there is one node for which both bits are set.

We can use a similar structure for the test between $S_R$ and $T_R$ in line 5.
$S_{R}$ and $T_{R}$ only grow as long as the true-branch in lines 6-8 is not executed.
Outside of the true-branch we can therefore use the same bit-vector trick to achieve an $O(1)$-test in line 5.
The lines 6-8 consist of the code that augments the flow, i.e., they have a running time of $O(m)$ each time that the branch is executed.
In $O(m)$ running time we can reset the bit-arrays, i.e., entering the true-branch is unproblematic for the running time of the test in line 5.
We can therefore account for the running time needed to manage the bit-arrays in the lines 6-8 and have an $O(1)$-test in line 5.

As already stated, the lines 6-8 augment the flow and need $O(m)$ running time each time that they are executed. 
Fortunately, there can be at most $c$ path augmentations.
The total time spent in the lines 6-8 over the algorithm's execution is therefore  in~$O(cm)$.

In addition to maintaining the bit-arrays for $S_R$ and $T_R$, we can keep track of the number of elements in the sets.
This allows us to implement the test in line 10 in $O(1)$.

Showing that the algorithm spends no more than $O(cm)$ running time in the lines 11-15 and in the analogous lines 16-17 is the tricky part of the algorithm's analysis. 
The lines 16-17 follow directly by symmetry and therefore we focus on the lines 11-15.

We will first establish that the lines 10-17 are only executed at most $m$ times.
In each iteration an arc is chosen as piercing arc.
After being chosen, an arc cannot participate in another cut and can therefore not be chosen a second time as piercing arc.
As there are only $m$ arcs, the number of iterations is bounded by $m$.

For each of $S$, $T$, $S_R$, and $T_R$ we maintain the data structures of a breadth-first search\footnote{A depth-first search would work too.}, i.e., a queue and a bit-array of $n$ elements. 
Growing a set as seen in the lines 11 and 15 consists of removing nodes from the corresponding queue and visiting neighboring nodes until the queue is emptied, i.e., executing the regular breadth-first search algorithm.
Adding a node to the set as seen in line 14 consists of adding the node to queue and raising the corresponding in the bit-array.
It is thus clear the operation in line 14 is in $O(1)$ and as there are at most $m$ iterations, the total time spent in line 14 is in $O(m)$ which is below the claimed running time of $O(cm)$.
The growing of the sets $S$ and $T$ is also in $O(m)$ as we never remove an element from $S$ nor $T$ and they therefore consist of standard breadth-first searches.
These searches are interrupted from time to time but this does not change the fact that the total running time spent in them is in $O(m)$.
Analyzing the running time required to grow the sets $S_R$ and $T_R$ is more difficult as the states of the associated searches can be reset in line 7.
Fortunately, as we have already established, line 7 can only be executed at most $c$ times.
There are therefore only $O(c)$ state resets.
Between two resets the search consists of a normal breadth-first-search with a running time in $O(m)$.
The total running time is therefore bounded by $O(cm)$.

We assumed that the piercing oracle requires a running time proportional to the number of arcs in the cut from which it must chose.
The number of cut arcs never decreases.
Further, there are $c$ cut arcs at the end.
We therefore know that $c$ is an upper bound to the size of every intermediate cut.
Further, as there are at most $m$ iterations, we have bounded the total running time in line 13 by $O(cm)$.

It remains to show that line 12, which outputs the cuts, does not require more than $O(cm)$ running time.
This seems trivial at first but the details are significantly more involved than one would naively expect.
Following the argumentation for line 13, we know that the operation must run in $O(c)$ running time to achieve a total running time of $O(cm)$.
The algorithm must therefore output the cuts as list of cut arcs and not as bit-array that maps each node to a side, as is often done in competitor algorithms.
Outputting bit-arrays would be too slow.
Another problem consists of identifying the cut-arcs efficiently.
In $O(c)$ running time, the algorithm cannot iterate over all nodes in $S$ or $T$ to determine all outgoing arcs, which is needed to find the cut-arcs in the straight-forward way.
The trick to achieve the required running time consists of maintaining two lists of saturated arcs.
The first list consists of saturated arcs that depart in $S$ and could be part of the cut.
The second list consists of saturated arcs that enter $T$ and works analogously.
If the algorithm encounters a saturated arc when growing $S$ in line 11, it adds the arc to the list of $S$.
It does this regardless of whether the arc is a cut arc.
When reaching line 12, this list contains all cut arcs but also possibly additional saturated arcs that are not part of the cut.
The algorithm therefore iterates over all arcs before outputting them and removes those that are no cut arcs.
This step can have a running time larger than $O(c)$.
Fortunately, as every arc can only be added once to the list, it can also only be removed once.
The total running time needed for the removal is therefore in $O(m)$ and we do not need to account for it in line 12.
Further, after removing superfluous arcs at most $O(c)$ remain, which is within the required bounds.
This concludes the proof that the running time of our core algorithm is within~$O(cm)$.

\subsection{Piercing Heuristic}
\label{sec:Pierce-Heuristic}

\begin{figure}
\begin{center}
\includegraphics[scale=1]{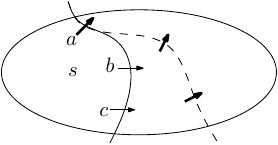}
\end{center}

\caption{\label{fig:avoid-path}The curves represent cuts, the current one is solid.
The arrows are cut-arcs, bold ones result in augmenting paths.
The dashed cut is the next cut where piercing any arc results in an augmenting path.}
\end{figure}%

In this section, we describe how we implement the piercing oracle used in the previous section. 
Given an unbalanced arc cut $C$, the piercing oracle should select a piercing arc that is not part of the final balanced cut in at most $O(\left|C\right|)$ time. 
Piercing the source side and target side cuts are analogous and we therefore only describe the procedure for the source side. 
Denote by $a=(q,p)$ the piercing arc with piercing node $p\not\in S$. 

Our piercing heuristic is composed of two parts: The primary and the secondary heuristic.
The primary heuristic first narrows down the set of potential piercing arcs and the secondary heuristic then chooses from this smaller set.

\subsection{Primary Heuristic: Avoid Augmenting Paths}

The first heuristic consists of avoiding augmenting paths whenever possible. 
Piercing an arc $a$ leads to an augmenting path, if and only if $p\in T_{R}$, i.e., a non-saturated path from $p$ to a target node exists. 
As our algorithm has computed $T_{R}$, it can determine in constant time whether piercing an arc would increase the size of the next cut. 
The proposed heuristic consists of preferring edges with $p\not\in T_{R}$ if possible. 
It is possible that none or multiple $p\not\in T_{R}$ exist. 
In this case our algorithm employs a further heuristic to choose the piercing arc among them. 

However, the secondary heuristic is often only relevant in the case that an augmenting path in unavoidable. 
Consider the situation depicted in Figure~\ref{fig:avoid-path}.
Our algorithm can choose between three piercing arcs $a$, $b$, and $c$. 
It will not pick $a$ as this would increase the cut size. 
The question that remains is whether $b$ or $c$ is better. 
The answer is that it nearly never matters. 
Piercing $b$ or $c$ does not modify the flow and therefore not $T_R$. 
Which piercing arcs result in larger cuts is therefore left unchanged. 
No matter whether $b$ or $c$ is picked, picking $a$ in the next iteration results again in an augmenting path.
The algorithm will therefore eventually end up with the same cut composed solely of arcs that should be avoided unless perfect balance is achieved first.
This cut is represented as dashed line in Figure~\ref{fig:avoid-path}.
We know that the dashed cut has the same size as all cuts found between the current cut and the dashed cut. 
Further, the dashed cut has the best balance among them and therefore dominates all of them. 
It therefore does not matter which of these dominated cuts are enumerated and in which order they are found.

This means that most of the time our avoid-augmenting-paths heuristic does the right thing. 
However, it is less effective when cuts approach perfect balance. 
In this case it is possible that perfect balance is achieved before the dashed cut is found.
The result consists of a race between source and target sides to claim the last nodes. 
Not the best side wins, but the first that gets there.

\subsection{Secondary Heuristic: Distance-Based}

If our primary avoid-augmenting-paths heuristic does not uniquely determine the piercing arc, we use a secondary distance heuristic to tie-break between the remaining choices.
Our algorithm picks a piercing arc such that $\dist(p,t)-\dist(s,p)$ is maximized, where $s$ and $t$ are the original source and target nodes. 
The $\dist(p,t)$-term avoids that the source side cut and the target side cut meet as nodes close to $t$ are more likely to be close to the target side cut.
Subtracting $\dist(s,p)$ is motivated by the observation that $s$ has a high likelihood of being positioned far away from the balanced cuts. 
A piercing node close to $s$ is therefore likely on the same side as $s$. 
Our algorithm precomputes the distances from $s$ and $t$ to all nodes before the core algorithm is run. 
This allows it to evaluate $\dist(p,t)-\dist(s,p)$ in constant time inside the piercing oracle. 

\begin{figure}
\begin{center}
\includegraphics[scale=0.75]{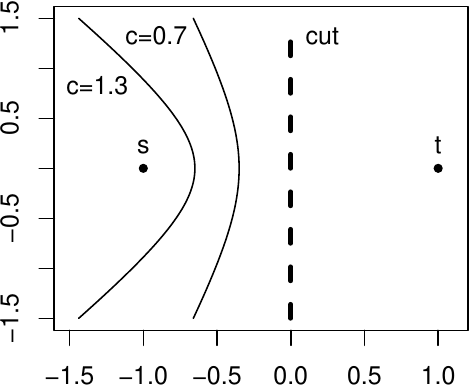}
\end{center}

\caption{\label{fig:dist-heuristic}Geometric interpretation of the distance
heuristic.}
\end{figure}

The distance heuristic has a geometric interpretation as depicted in Figure~\ref{fig:dist-heuristic}. 
We interpret the nodes as positions in the plane and the distances as being euclidean. 
The set of points $p$ for which $\left\Vert p-t\right\Vert _{2}-\left\Vert p-s\right\Vert _{2}=c$ holds for some constant $c$ is one branch of a hyperbola whose two foci are $s$ and $t$. 
The figure depicts the branches for $c=1.3$ and $c=0.7$. 
The heuristic prefers piecing nodes on the $c=1.3$-branch as it maximizes $c$. 
A consequence of this is that the heuristic works well if the desired cut follows roughly a line perpendicular to the line through $s$ and $t$.
This heuristic works on many graphs but there are instances where it breaks down.
For example cuts with a circle-like shape are problematic.
This geometric interpretation also works in higher-dimensional spaces.

\section{Extensions}
\label{sec:Extensions}

Our base algorithm can be extended to compute general small cuts that are independent of an input $st$-pair, to compute node separators, and to compute contraction orders.

\subsection{General Cuts}

Our core algorithm computes balanced $st$-cuts. 
In many situations cuts independent of a specific $st$-pair are needed. 
This problem variant can be solved with high probability by running FlowCutter $q$ times with $st$-pairs picked uniformly at random. 
Indeed, suppose that $C$ is a Pareto-optimal cut such that the larger side has $\alpha\cdot n$ nodes (i.e. $\alpha=(\epsilon+1)/2$) and $q$ is the number of $st$-pairs.
The probability that $C$ separates a random $st$-pair is $2\alpha(1-\alpha)$.
The success probability over all $q$ $st$-pairs is thus $1-\left(1-2\alpha(1-\alpha)\right)^{q}$.
For $\epsilon=33\%$ and $q=20$, the number of pairs we recommend in our experiments, the success probability is larger than 99.99\%. 
For larger $\alpha$ this rate decreases. 
However, it is still large enough for all practical purposes, as for $\alpha=0.9$ (i.e. $\epsilon=80\%$) and $q=20$ the rate is still slightly above 98.11\%. 
The number of $st$-pairs needed does not depend on the size of the graph nor on the cut size. 
If the instances are run one after another then the running time depends on the worst cut's size which may be more than $c$.
We therefore run the instances simultaneously and stop once one instance has found a cut of size $c$.
The running time is thus in $O(qcm)$.
As we set $q$ to a constant value of at most 100 in our experiments, the running time is in $O(cm)$.

This argumentation relies on the assumption that it is enough to find an $st$-pair that is separated.
However, in practice the positions of $s$ and $t$ in their respective sides influence the performance of our piercing heuristic. 
As a result it is possible that in practice more $st$-pairs are needed than predicted by the argument above because of effects induced by the properties of the piercing oracle.

\subsection{\label{sec:node-sep}Node Separators}

\begin{figure}
\begin{center}
\subfloat[Input graph]{\begin{centering}
\includegraphics{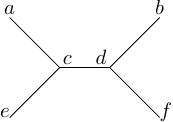}
\end{centering}

}~~~~\subfloat[Expanded graph]{\begin{centering}
\includegraphics{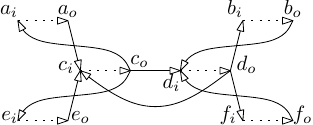}
\end{centering}

}
\end{center}

\caption{\label{fig:expansion}Expansion of an undirected graph $G$ into a
directed graph $G'$. The dotted arrows are internal arcs. The solid
arrows are external arcs. }
\end{figure}

To compute contraction orders, node separators are needed and not edge cuts.
To achieve this, we employ a standard construction to model node capacities in flow problems~\cite{ff-fn-62,amo-nf-93}. 
We transform the symmetric input graph $G=(V,A)$ into a directed expanded graph $G'=(V',A')$ and compute flows on $G'$. 
We expand $G$ into $G'$ as follows: For each node $x\in V$ there are two nodes $x_{i}$ and $x_{o}$ in $V'$.
We refer to $x_{i}$ as the \emph{in-node} and to $x_{o}$ as the \emph{out-node} of $x$. 
There is an \emph{internal arc} $(x_{i},x_{o})\in A'$ for every node $x\in V$. 
We further add for every arc $(x,y)\in A$ an \emph{external arc} $(x_{o},y_{i})$ to $A'$. 
The construction is illustrated in Figure~\ref{fig:expansion}. 
For a source-target pair $s$ and $t$ in $G$ we run the core algorithm with source node $s_{o}$ and target node $t_{i}$ in $G'$. 
The algorithm computes a sequence of cuts in $G'$. 
Each of the cut arcs in $G'$ corresponds to a separator node or a cut edge in $G$ depending on whether the arc in $G'$ is internal or external. 
From this mixed cut our algorithm derives a node separator by choosing for every cut edge in $G$ the endpoint on the larger side. 
Unfortunately, using this construction, it is possible that the graph is separated into more than two components, i.e., we can no longer guarantee that both sides are connected.

\subsection{Contraction Orders}
\label{sec:order}

\begin{figure}

\begin{center}
\includegraphics[scale=0.75]{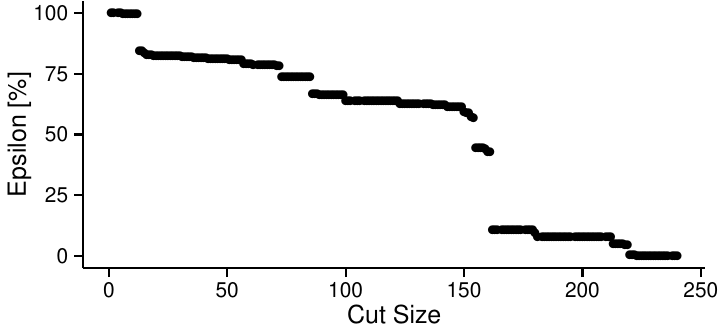}
\end{center}

\caption{\label{fig:pareto-plot}Edges cuts found by FlowCutter with 20 random source-target pairs for the Central Europe graph used in the experiments.}
\end{figure}

Our algorithm constructs contraction orders using an approach based on nested dissection \cite{g-ndrfe-73,lrt-gnd-79}. 
It bisects $G$ along a node separator $Q$ into subgraphs $G_{1}$ and $G_{2}$. 
It recursively computes orders for $G_{1}$ and $G_{2}$. 
The order of $G$ is the order of $G_{1}$ followed by the order of $G_{2}$ followed by the nodes in $Q$ in an arbitrary order. 
Selecting $Q$ is unfortunately highly non-trivial.

The cuts produced by FlowCutter can be represented using a plot such as in Figure~\ref{fig:pareto-plot}.
Each point represents a non-dominated cut.
The question is which of the many points to choose.
After some experimentation, we went with the following heuristic: 
Pick a separator with minimum expansion and at most 60\% imbalance. 
This choice is not perfect as the experiments of Section~\ref{sec:europe-graph} show but works well in most cases.
Picking a cut of minimum expansion given a Pareto-cut-set is easy.
However, we know of no easy way to do this using an algorithm that computes a single small cut of bounded imbalance, as all the competitor algorithms do.
It is therefore not easy to swap FlowCutter out for any of the competitors without also dropping expansion as optimization criterion.

We continue the recursion until we obtain trees and cliques. 
On cliques any order is optimal.
On trees an order can be derived from a so called optimal node ranking as introduced in~\cite{irv-o-88}.
A node ranking of a tree is a labeling of the nodes with integers $1,2\ldots k$ such that on the unique path between two nodes $x$ and $y$ with the same label there exists a node $z$ with a larger label.
An optimal node ranking is one with minimum $k$.
Contracting the nodes by increasing label yields an elimination tree of minimum depth.
In~\cite{s-o-89} it has been shown that these ranking can be computed in linear running time. 

\paragraph{Special Preprocessing for Road Graphs}

Road graphs have many nodes of degree 1 or 2. 
We exploit this in a fast preprocessing step similar to \cite{dsw-fespd-sig15} to significantly reduce the graph size. 

Our algorithm first determines the largest biconnected component $B$ using \cite{ht-eagm-73} in linear time. 
It then removes all edges from $G$ that leave $B$. 
It continues independently on every connected component of $G$ as described in the next paragraph. 
The set of connected components usually consists of $B$ and many tiny often tree-like graphs.
The resulting orders are concatenated such that the order of $B$ is at the end. 
The other orders can be concatenated arbitrarily.
 
For each connected component our algorithm starts by marking the nodes with a degree of 3 or higher.
For each degree-2-chain $x,y_{1}\ldots y_{k},z$ between two marked nodes $x$ and $z$ with $x\neq z$, it adds an edge $\{x,z\}$.
It then splits the graph into the graph $G_{\ge 3}$ induced by the marked nodes and the graph $G_{\le 2}$ induced by the unmarked nodes.
Edges between marked and unmarked nodes are dropped.
$G_{\le 2}$ consists of the disjoint union of paths.
As paths are trees, we can therefore employ the node-ranking-based tree-ordering algorithm described above to determine a node order.
For $G_{\ge 3}$ we determine an order using FlowCutter and nested dissection as described above
We position the nodes of $G_{\le 2}$ before those of $G_{\ge 3}$ and obtain the node order of the connected component.

\section{Experiments}
\label{sec:experiments}

We compare Flowcutter to the state-of-the-art partitioners KaHip \cite{ss-tlagh-13}, Metis \cite{kk-mspig-99}, and InertialFlow \cite{ss-obsrn-15}.
There is a superficial comparison with PUNCH~\cite{dgrw-gpnc-11} in Section \ref{sec:punch}.
We present three experiments: (1) we compare the produced contraction orders in terms of CCH performance in Section~\ref{sec:order-experiments} on road graphs made available during the DIMACS challenge on shortest paths~\cite{dgj-spndi-09}, (2) compare the Pareto-cut-sets in more detail in Section~\ref{sec:experiments-pareto} on the same road graphs, and (3) evaluate FlowCutter on non-road graphs using the Walshaw benchmark set~\cite{swc-acesm-04} in Section~\ref{app:walshaw}. 
Section \ref{sec:algo-setup} describes the experimental setup common to all experiments.
All experiments were run on a Xeon E5-1630 v3 @ 3.70GHz with 128GB DDR4-2133 RAM. 

\subsection{Algorithm Implementations Used and Their Configurations}
\label{sec:algo-setup}

\paragraph{Edge Cut Algorithm}
We use FlowCutter in three variants denoted by F3, F20, and F100, with 3, 20, and 100 random source-target-pairs respectively. 
InertialFlow was introduced in~\cite{ss-obsrn-15} but no code was published. 
Fortunately, the idea is simple and we therefore were able to reimplement the algorithm.
We refer to it using the letter I.
Metis is a well-known general graph partitioner based on a multi-level scheme.
The original authors published source code, which we used in our experiments.
We compare against Metis 5.1.0 which is the newest version at the time of writing and refer to it as M.
KaHip also uses a multi-level scheme but adds a lot of optimizations compared to Metis that can drastically decrease the cut sizes.
The KaHip source code is also available and thus we use it for our experiments.
At the time of writing, the current version of KaHip is 1.00.
Unfortunately, we have observed several regressions compared to earlier versions.
These regressions are due to a bug being fixed that caused certain expensive flow-based refinement steps not being run for higher imbalances. 
The newer version achieves smaller cuts at the expense of higher running times for higher imbalances.
Because of these regressions and for comparability with previous works, we also include comparisons with the earlier versions KaHip 0.61 and KaHip 0.73 which were the current versions and therefore used when we performed the experiments for~\cite{dsw-cch-15} and~\cite{hs-gbpo-16} respectively.
We use KaHip in the strong preconfiguration and add \texttt{-{}-enforce\_balance} to the commandline for $\max \epsilon = 0$.
We refer to the three variants as K0.61, K0.73, and K1.00.

\paragraph{Node Ordering Algorithms}
Metis provides its own node ordering tool called \texttt{ndmetis}, which we use.
Unfortunately, no other package provides a similar tool.
We have therefore implemented a nested dissection algorithm on top of them. 
For KaHip 1.00 and InertialFlow we use the same straight-forward nested dissection implementation that computes one edge cut at each level and recurses until either cliques or trees are reached. 
Edge cuts are transformed into node separators by picking the nodes on one side incident to the cut edges.
For KaHip we use a maximum imbalance of 20\% and for InertialFlow we use 60\%.
For KaHip 0.61 we use an older nested dissection implementation originally written for~\cite{dsw-cch-15}.
It is not optimized for running time and only for quality.
At each level, it invokes KaHip several times with different random seeds and picks the smallest cut found.
It calls KaHip repeatedly on every level until for ten consecutive calls no smaller cut is found.
We do this to reliably get rid of variations in achieved cut sizes that are due to randomization.
However, this setup is unfavorable to KaHip as it results in large running times.
We decided to stick with the old ordering routine for K0.61 for comparability with~\cite{dsw-cch-15} and use an ordering scheme for K1.0 that only computes one cut per level.
The FlowCutter nested dissection implementation is based on the same code as used for KaHip 1.00 and InertialFlow but uses the separator variant of FlowCutter and performs the low-degree node optimizations described in Section~\ref{sec:order}.

KaHip v1.00 includes a more sophisticated tool to transform edge cuts into node separators using the algorithm of~\cite{ss-amnsa-16}. 
We tried using it in combination with the newer nested dissection scheme with one separator per level, but we needed 19 hours to compute orders for the small California and Nevada graph used in our experiments. 
We were not able to compute orders on the larger instances in reasonable time and therefore omit this algorithm from our comparison.

\subsection{Order Experiments}
\label{sec:order-experiments}

We computed contraction orders for 4 DIMACS roads graphs~\cite{dgj-spndi-09}. Our results are summarized in Table~\ref{tab:Contraction-Orders}.

\paragraph{Instances}

The smallest test instance is the DIMACS Colorado graph with $n=436\mathrm{K}$ and $m=1\mathrm{M}$. 
Next is California and Nevada with $n=1.9\mathrm{M}$ and $m=4.6\mathrm{M}$, followed by (Western) Europe with $n=18\mathrm{M}$ and $m=44\mathrm{M}$ and finally a graph encompassing the whole USA with $n=24\mathrm{M}$ and $m=57\mathrm{M}$.

\begin{table}

\setlength\tabcolsep{3.0pt}

\begin{center}
\begin{tabular}{ccrrrrrrrrrr}
\toprule
 &  & \multicolumn{4}{c}{Search Space} & \#Arcs  &  & Up. & \multicolumn{3}{c}{Running times}  \\
\cmidrule(lr){3-6} \cmidrule(lr){10-12}
 &  & \multicolumn{2}{c}{Nodes} & \multicolumn{2}{c}{Arcs {[}$\cdot10^{3}${]}} & in CCH & \#Tri. & Tw. & Order & Cust. & Query\\
\cmidrule(lr){3-4} \cmidrule(lr){5-6}   
 &  & Avg. & Max. & Avg. & Max. & {[}$\cdot10^{6}${]} & {[}$\cdot10^{6}${]} & Bd. & {[}s{]} & {[}ms{]} & {[}$\mu$s{]}\\
\midrule
\multirow{7}{*}{\begin{sideways}
Col
\end{sideways}} & M & 155.6 & 354 & 6.1 & 22 & 1.4 & 6.4 & 102 & \textbf{2.0} & 18 & 26\\
 & K0.61 & 135.1 & 357 & 4.6 & 22 & 1.7 & 7.2 & 103 & 3\,837.1 & 21 & 20\\
 & K1.00 & 136.4 & 357 & 4.8 & 22 & 1.5 & 6.9 & 99 & 1\,052.4 & 20 & 20\\
 & I & 151.2 & 542 & 6.2 & 38 & 1.5 & 7.4 & 119 & 7.4 & 21 & 24\\
 & F3 & 126.3 & 280 & 4.1 & 15 & 1.3 & 4.8 & 91 & 10.3 & 15 & 18\\
 & F20 & \textbf{122.4} & \textbf{262} & \textbf{3.8} & \textbf{14} & \textbf{1.3} & \textbf{4.4} & \textbf{87} & 61.0 & \textbf{14} & \textbf{17} \\
 & F100 & 122.5 & 264 & \textbf{3.8} & \textbf{14} & \textbf{1.3} & \textbf{4.4} & \textbf{87} & 285.9 & \textbf{14} & 18 \\
\midrule
\multirow{7}{*}{\begin{sideways}
Cal
\end{sideways}} & M & 275.5 & 543 & 17.3 & 53 & 6.5 & 36.4 & 180 & \textbf{9.9} & 88 & 60\\
 & K0.61 & 187.7 & 483 & 7.0 & 37 & 7.5 & 34.2 & 160 & 18\,659.3 & 89 & 30\\
 & K1.00 & 184.9 & 471 & 6.8 & 38 & 7.0 & 33.4 & 143 & 6\,023.6 & 86 & 30\\
 & I & 191.4 & 605 & 7.1 & 53 & 6.9 & 34.1 & 161 & 42.6 & 84 & 31\\
 & F3 & 177.5 & \textbf{356} & 6.2 & \textbf{24} & 5.9 & 23.4 & \textbf{127} & 64.1 & 69 & 27\\
 & F20 & 170.0 & 380 & \textbf{5.6} & 26 & \textbf{5.8} & \textbf{21.8} & 132 & 386.8 & 68 & \textbf{26}\\
 & F100 & \textbf{169.5} & 380 & \textbf{5.6} & 26 & \textbf{5.8} & \textbf{21.8} & 132 & 1\,831.8 & \textbf{65} & \textbf{26}\\
\midrule
\multirow{7}{*}{\begin{sideways}
Eur
\end{sideways}} & M & 1\,223.4 & 1\,983 & 441.4 & 933 & 69.9 & 1\,390.4 & 926 & \textbf{125.9} & 2\,241 & 1\,164\\
 & K0.61 & 638.6 & 1\,224 & 114.3 & 284 & 73.9 & 578.2 & 482 & 213\,091.1 & 971 & 303\\
 & K1.00 & 652.5 & 1\,279 & 113.4 & 287 & 68.3 & 574.5 & 451 & 242\,680.5 & 934 & 297\\
 & I & 732.9 & 1\,569 & 149.7 & 414 & 67.4 & 589.7 & 516 & 1\,017.2 & 935 & 385\\
 & F3 & 734.1 & 1\,159 & 140.2 & 312 & 60.3 & 519.4 & 531 & 2\,531.6 & 853 & 365\\
 & F20 & \textbf{616.0} & \textbf{1\,102} & \textbf{102.8} & 268 & \textbf{58.8} & 459.6 & 455 & 16\,841.5 & 784 & \textbf{270}\\
 & F100 & 622.6 & 1\,105 & 104.8 & \textbf{239} & \textbf{58.8} & \textbf{459.4} & \textbf{449} & 85\,312.8 & \textbf{766} & 278\\
\midrule
\multirow{7}{*}{\begin{sideways}
USA
\end{sideways}} & M & 990.9 & 1\,685 & 249.1 & 633 & 86.0 & 1\,241.1 & 676 & \textbf{170.8} & 2\,111 & 651\\
 & K0.61 & 575.5 & 1\,041 & 71.3 & 185 & 97.9 & 737.1 & 366 & 265\,567.3 & 1\,250 & 202\\
 & K1.00 & 540.3 & 1\,063 & 62.3 & 208 & 88.7 & 648.3 & 439 & 315\,942.6 & 1\,097 & 179\\
 & I & 533.6 & 1\,371 & 62.0 & 291 & 88.8 & 682.0 & 384 & 1\,076.8 & 1\,125 & 177\\
 & F3 & 562.7 & 906 & 66.4 & 159 & 75.9 & 478.4 & 321 & 2\,108.7 & 858 & 191\\
 & F20 & \textbf{490.6} & 868 & \textbf{52.7} & \textbf{154} & \textbf{74.3} & \textbf{440.5} & 312 & 12\,379.2 & \textbf{812} & 156\\
 & F100 & 490.9 & \textbf{863} & 52.8 & \textbf{154} & \textbf{74.3} & 442.6 & \textbf{311} & 59\,744.6 & 886 & \textbf{155} \\
\bottomrule
\end{tabular}
\end{center}

\caption{\label{tab:Contraction-Orders}Contraction Order Experiments. 
We report the average and maximum over all nodes~$v$ of the number of nodes and arcs in the CCH-search space of $v$, the number of arcs and triangles in the CCH, and the induced upper treewidth bound. 
We additionally report the order computation times, the customization times, and the average shortest path distance query times. 
Only the customization times are parallelized using four cores. 
The customization times are the median over nine runs to eliminate variance. The query running times are averaged over $10^6$ $st$-queries with $s$ and $t$ picked uniformly at random.
Several CCH customization variants exist. The times reported are for a non-amortized, non-perfect customization, with SSE and uses precomputed triangles.
}

\end{table}

\paragraph{Relations between Columns}
Table~\ref{tab:Contraction-Orders} contains a lot of data.
However, some columns are related.
We therefore first point these relations out and then limit our discussion to the remaining non-related columns.
We observe that, modulo small cache effects, the customization time is correlated with the number of triangles and the average query running time is correlated with the number of arcs in the CCH.
These correlations are non-surprising and were predicted by CCH theory.
Denote by $n_s$ and $m_s$ the number of nodes and arcs in the search space. 
For the average numbers we observe that $1.7\le \frac{n_s(n_s-1)}{2} / m_s \le 2.6$ and for the maximum numbers we observe that $2.1 \le \frac{n_s(n_s-1)}{2} / m_s \le 3.9$, which indicates that the search spaces are nearly complete graphs.
The number of nodes and the number of arcs are thus related.
We can therefore say that search space is small or large without indicating whether we refer to nodes or arcs as one implies the other.

\paragraph{Search spaces}
One of the FlowCutter variants always produces the smallest search spaces.
KaHip produces the next smaller search spaces, followed by InertialFlow.
Metis is last by a large margin.
It is interesting that the USA graph has a smaller search space than the Europe graph.
The ratio between the average and the maximum search space sizes is very interesting. 
A high ratio indicates that a partitioner often finds good cuts, but at least one cut is comparatively bad.
This ratio is never close to 1, indicating that road graphs are not perfectly homogeneous.
In some regions, probably cities, the cuts are worse than in some other regions, probably the country-side.
However, compared to the competitors, the ratio is higher for InertialFlow.
This illustrates that its geography-based heuristic is effective most of the time but in few cases fails noticeably at finding a good cut. 

\paragraph{Number of Arcs}
A small search space size is not equivalent with the CCH containing only few arcs.
It is possible that vertices are shared between many search spaces and thus the CCH can be significantly smaller than the sum of the search space sizes.
This effect occurs and explains why the number of arcs in CCH is orders of magnitude smaller than the sum over the arcs in all search spaces.
Further, minimizing the number of arcs in the CCH is not necessarily the same as minimizing the search space sizes.
This explains why Metis beats KaHip in terms of CCH size but not in terms of search space size.
InertialFlow seems to be comparable to Metis in terms of CCH size, as their CCH arc counts are never significantly different.
However, FlowCutter beats all competitors and clearly achieves the smallest CCH sizes.

\paragraph{Number of Triangles}
A third important order quality metric is the number of triangles in the CCH.
Metis is competitive on the two smaller graphs, but is clearly dominated on the continental sized graphs.
InertialFlow and KaHip seem to be very similar on all but the USA graph.
On the USA graph K1.0 is ahead of both InertialFlow and K0.61.
FlowCutter also wins with respect to this quality metric producing between 20\% and 30\% fewer triangles than the closest competitor.

\paragraph{Treewidth}
As the CCH is essentially a chordal graph which are closely tied to tree-decompositions, we can easily obtain upper bounds on the tree-width of the input graphs as a side product.
This quality metric is not directly related to CCH performance, but is of course indirectly related as most of the other criteria can be bounded in terms of it.
As such it reflects the same trend: Metis is worst, followed by InertialFlow, followed by KaHip, and FlowCutter with the best bounds.
Analogous to the search space sizes, we observe that the USA graph has a significantly lower tree-width than the Europe graph, assuming that our upper treewidth bounds are not completely off.

\paragraph{Running Time}
Quality comes at a price and thus the computation times of the orders follow nearly the opposite trend: KaHip is the slowest, followed by FlowCutter, followed by InertialFlow, while Metis is astonishingly fast.

\paragraph{K1.00 vs K0.61}
The two KaHip versions seem to be very similar.
Sometimes the newer version K1.00 is ahead and sometimes the older version K0.61 wins in terms of order quality.
We explain this effect by differences in implementation in our nested dissection code.
Recall that K0.61 takes the best cut of at least 10 iterations on each level, whereas K1.00 only computes a single cut. 
This means that K1.00 is more sensible to random fluctuations coming from bad random seeds than K0.61.
On average, one run of K1.00 is better than one run of K0.61. 
However, the best of at least 10 K0.61 runs wins against one K1.00 run with a bad seed.
This effect explains the observed variance.
Both, the running times of K1.0 and K0.61, are very high but for different reasons.
K0.61 is slow because of the numerous repetitions on each level.
However, K1.00 is slow because the newer KaHip version is significantly slower for $\epsilon=20\%$ than the older versions.
We will see this effect in greater detail in Section~\ref{sec:experiments-pareto}.

\paragraph{F3 vs F20 vs F100}
It is not always clear which of F3, F20, or F100 gives the best results.
F3 is most of the time slightly worse. 
This suggests that three source-target pairs are enough to get good separators most of the time but not enough to be fully reliable. 
A bad random seed can result in good separators being missed.
The difference between F20 and F100 in terms of order quality is nearly negligible. 
This means that F20 and F100 find nearly always at least very similar separators.
We conclude that there is no real advantage of going from 20 source-target pairs to 100 on road graphs.
20 source-target pairs are enough to be quality wise nearly independent of the random seed used.

\subsection{Pareto Cut Set Experiments}
\label{sec:experiments-pareto}

\newcommand{\makeparetotable}[2]{

\begin{table}

\setlength\tabcolsep{3pt}
\setlength\mycolwidth{1.0cm}
\setlength\mysmallcolwidth{0.8cm}
\begin{center}

\begin{tabular}{rR{\mycolwidth}R{\mycolwidth}R{\mycolwidth}R{\mycolwidth}R{\mycolwidth}R{\mycolwidth}R{\mysmallcolwidth}R{\mysmallcolwidth}R{\mysmallcolwidth}R{\mysmallcolwidth}R{\mysmallcolwidth}R{\mysmallcolwidth}}

\toprule

\multirow{2}{*}{\rotatebox[origin=c]{90}{$\max\epsilon$}}  & \multicolumn{6}{c}{Achieved $\epsilon$ [\%]} & \multicolumn{6}{c}{Cut Size}\\
\cmidrule(lr){2-7}  \cmidrule(lr){8-13} 
   & F3 & F20 & K0.73 & K1.00 & M & I & F3 & F20 & K0.73 & K1.00 & M & I\\
\midrule

\csvreader[late after line=\\,late after last line=\\\midrule]{#1.first_row.out.csv}{}%
{\csvcoli & \csvcolii & \csvcoliii & \csvcoliv & \csvcolv & \csvcolvi & \csvcolvii & \csvcolviii & \csvcolix & \csvcolx & \csvcolxi & \csvcolxii & \csvcolxiii}%

\multirow{2}{*}{\rotatebox[origin=c]{90}{$\max\epsilon$}} & \multicolumn{6}{c}{Running Time [s]} & \multicolumn{6}{c}{Are sides connected?}\\
\cmidrule(lr){2-7}  \cmidrule(lr){8-13} 
 & F3 & F20 & K0.73 & K1.00 & M & I & F3 & F20 & K0.73 & K1.00 & M & I\\
\midrule

\csvreader[late after line=\\,late after last line=\\\bottomrule]{#1.second_row.out.csv}{}%
{\csvcoli & \csvcolii & \csvcoliii & \csvcoliv & \csvcolv & \csvcolvi & \csvcolvii & \csvcolviii & \csvcolix & \csvcolx & \csvcolxi & \csvcolxii & \csvcolxiii}%

\end{tabular}
\end{center}

\caption{#2}
\end{table}

}

\makeparetotable{pareto/usa}{\label{tab:pareto-usa}Results for the DIMACS USA graph.}

In the previous experiment, we have demonstrated that FlowCutter produces the best contraction orders.
In this section, we look at the Parteo-cut sets of five graphs in more detail.
These are the DIMACS California and Nevada, Colorado, USA, and Europe graphs and a Central European subgraph.

\paragraph{Experimental Setup}

For each of these graphs we report the results in a table similar to Table~\ref{tab:pareto-usa}.
With the exception of FlowCutter, we ran each of the algorithms for various maximum imbalance input parameters (max $\epsilon$ column), effectively sampling the Pareto-set computed by each partitioner.
We report the imbalance of the produced cut.
This achieved imbalance can be smaller than the input parameter which is only a maximum.
We further report the size of each cut and indicate whether both sides of the cut form connected subgraphs.
Finally, we report the running time needed to compute each cut.
To compute all reported cuts, i.e., the sampled Pareto-set, all partitioners except FlowCutter need the sum over all reported running times.

For FlowCutter we use a slightly different setup.
We compute a set of Pareto-cuts using FlowCutter and then pick the best cut from this set that has an imbalance below the requested maximum. 
This means that for FlowCutter one can compute all reported cuts within the time needed to compute the cut for the input parameter max $\epsilon$ = 0.

\paragraph{PUNCH}

\label{sec:punch}

In~\cite{dgrw-gpnc-11}, a competing algorithm named PUNCH was introduced.
Unfortunately, we do not have access to an implementation of it.
We therefore cannot perform experiments with this algorithm.
However, for the USA and Europe graphs, the original authors \cite{dgrw-gpnc-11} report cut sizes for an imbalance of 3\%. 
In their experimental setup, PUNCH is run 100 times with varying random seeds.
On average, PUNCH finds a cut with 130 edges for Europe and 70 edges for the USA.
The minimum cut size over 100 runs is 129 respectively 69 edges. 
K1.0 finds a cut with 130 edges for Europe graph and 69 edges for the USA.
We conclude, that the performance of PUNCH is comparable in terms of quality to K1.0.

\paragraph{Instance Selection}

Selecting meaningful and representative testing instances is difficult as can be seen from Table~\ref{tab:pareto-usa}.
For the imbalance between 20\% and 50\% all partitioners with the exception of Metis find a cut of the same size.
One can argue that this imbalance range is the most relevant for our application.
It is therefore hard to argue, based on this experiment, whether one partitioner is better than another in terms of cut quality because they are all quasi the same.
All cuts with 61 edges divide the USA along the Mississippi river into east and west.
This cut is so pronounced that nearly all partitioners manage to find it.
However, we cannot conclude from this experiment that all partitioners are interchangeable in terms of quality.
This experiment only illustrates that the USA graph is in some sense an easy instance and therefore not a good testing instance.
We therefore need to look at subgraphs of the USA to be able to observe the differences in quality, that definitely exist given the difference in contraction order qualities.
We provide results for the DIMACS California and Nevada graph and the DIMACS Colorado graph in Tables~\ref{tab:pareto-cal} and~\ref{tab:pareto-col}.
We also ran experiments on the DIMACS Europe graph.
However, because of the special geographical topology of Europe, which we discuss in detail in Section~\ref{sec:europe-graph}, this graph is also non-representative.
We therefore evaluate the algorithms on a Central European subgraph induced by nodes with a latitude $\in[45,52]$ and a longitude $\in[-2,11]$.
This subgraph has about $n=7\mathrm{M}$ nodes and $m=18\mathrm{M}$ arcs.

\subsubsection{Discussion for USA}

As already outlined, we cannot deduce much from the experimental results for the USA graph.
However, there are a few observations that are interesting nonetheless.
Most of these observations are also valid for all other test instance.
We will therefore refrain from repeating these observations when discussing the other graphs.

\paragraph{Limitations of Metis}

Metis is clearly dominated as it is the only partitioner unable to find the Mississippi.
We can further observe that for imbalances of 70\% and above Metis finds huge cuts.
This is most likely a bug in the implementation. 
Further, while Metis does find a highly balanced cut, it is not perfectly balanced and therefore formally not a valid output for the case $\max \epsilon = 0$.  

\paragraph{Limitations of KaHip}

The running times of K0.73 are comparatively small for imbalances of 20\% and higher.
This is not the behavior that one would expect from the algorithm description.
The running time is expected to grow with increasing imbalance as it does for K1.00.
The reason for this behavior is the bug that was fixed in version 1.00. 
Before this version, KaHip would not do the flow based refinement steps correctly.
KaHip was therefore faster but the achieved cuts can be very strange.
This fixed bug is also the reason why computing contraction orders with K1.00 is so slow.

\paragraph{Different Mississippi cuts}

Another interesting observation is that while nearly all partitioners are able to find a Mississippi cut, they find different variants of it.
All cuts have size 61 but the achieved imbalances vary. 
FlowCutter finds slightly smaller imbalances than KaHip and InertialFlow. 
The cuts found by FlowCutter are therefore marginally better.

\paragraph{Connected Sides}

FlowCutter guarantees by construction that both sides of each reported cut are connected.
The other partitioners give no such guarantees. 
This means that the exact problem variants that they solve are slightly different.
We therefore report for each of the other partitioners whether the cut they find happens to have connected sides.
It is interesting that this is nearly always the case. 
One of the exceptions is for example the 3\% imbalance of K1.00 with 69 edges.
This cut is also the only situation where FlowCutter is outperformed in terms of cut size on the USA graph.
However, the sides of this cut are not connected.
The cut is therefore not a valid solution with respect to the exact problem setting solved by FlowCutter.
This explains why it is not found.

\paragraph{Perfectly balanced Cuts}

Even though it is not useful for our application, it is interesting to compare the algorithms in terms of perfectly balanced cuts.
This is the case when $\max \epsilon = 0$ or formulated differently: The number of nodes on each side must not differ by more than one node.
Past research has partially focused on this special case. 
KaHip even includes a special postprocessing step called cycle-refinement to reduce the sizes of perfectly balanced cuts~\cite{ss-tlagh-13}.
The results are surprising.
Metis is not able to find perfectly balanced cuts as the balance of the achieved cut is larger than required. 
For this border case the achieved cuts are thus formally not valid.
Even though KaHip includes special code, the achieved cut sizes are large.
They even rival those of InertialFlow, a heuristic that in the case of perfect balance degenerates to sorting the nodes by longitude and cutting along the median.
KaHip's cycle-refinement clearly does not work on this kind of graph.
Even though FlowCutter was not designed to compute perfectly balanced cuts, it is capable of doing so.
Further these cuts found turn out to be that smallest among all competitors by a significant margin.

\subsubsection{Discussion for California and Nevada}

\makeparetotable{pareto/cal}{\label{tab:pareto-cal}Results for the DIMACS California and Nevada.}

\paragraph{Perfectly balanced cuts}

We include the DIMACS California \& Nevada graph in our benchmark because~\cite{dfgrw-a-14} were able to determine the optimal size of a perfectly balanced cut for this graph.
The optimum is 32 edges.
The best cut found by the partitioners evaluated in Table~\ref{tab:pareto-cal} contains 39 edges and was found by FlowCutter.
It is therefore off by 7 edges.
However, even with a slight imbalance, i.e., $\max \epsilon = 1\%$, F3, F20, and K0.73 are able to find a cut with 31 edges. 
As this cut is smaller than the smallest balanced cut, it is possible that this 31 edge cut is optimal.

\paragraph{Cut sizes}

The sizes of the cuts on California seem to be similar to those of the USA graph.
There is one small and very pronounced cut, the one with 29 edges, which is found by all partitioniers.
However, F20 is able to find a 28 and 24 edge cut for higher imbalances.
KaHip misses these cuts and sticks with the 29 edge cut.
It is also interesting that InertialFlow is able to find a good 29 edge cut with 2.7\% imbalance. 
Unfortunately, it does not find it when the input parameter is at $\max \epsilon = 3\%$ but at $\max \epsilon = 20\%$.
This means InertialFlow is capable of finding good cuts, but $\max \epsilon$ parameters that significantly differ from the desired $\epsilon$ have to be tried.

\subsubsection{Discussion for Colorado}

\makeparetotable{pareto/col}{\label{tab:pareto-col}Results for the DIMACS Colorado.}

\paragraph{Perfectly balanced cuts}

The authors of~\cite{dfgrw-a-14} were also able to determine the minimum size of a perfectly balanced cut on the Colorado graph. 
It has 29 edges.
While FlowCutter comes closest among all the evaluated partitioners, the cut found is again significantly larger by 8 edges.
For imbalances in the range of 1\% to 3\% F20, K0.73, and K1.00 manage to achieve cut sizes of 29 edges but no cut is perfectly balanced.
All of them are therefore suboptimal. 
For $\epsilon=5\%$ cuts smaller than 29 edges are found.

\paragraph{Cut Sizes}

In contrast to the USA graph, we observe different cut sizes for the different partitioners on this instance for the relevant imbalances.
We can therefore better deduce from this experiment whether a partitioner is better than another for our specific application.
We observe that F20 wins with respect to every imbalance except for $\max \epsilon = 3\%$ and $\max \epsilon = 5\%$ where K1.00 and K0.73 respectively win by one edge.
This demonstrates that FlowCutter is indeed a heuristic and does not always achieve the optimum.
Comparing K1.00 and K0.73 is interesting.
One could expect K1.00 to always win because it is the newer version but this is not the case.
For $\max \epsilon = 1\%$ K1.00 is 5 edges ahead but for $\max \epsilon = 10\%$ K0.73 wins by 5 edges.
This can mean that K1.00 is not always superior to K0.73.
Another explanation is that both do not make enough iterations in their standard configuration to produce results that are reliable, i.e, with high probability insensitive to the random seed used.
Rerunning K1.00 and K0.73 with different random seeds could change the outcome.
The cut sizes of Metis are far from the competitors.
InertialFlow is better than Metis but also clearly dominated. 

\paragraph{Running Times}

Metis and InertialFlow are by an order of magnitude faster but also compute worse cuts.
The comparison between FlowCutter and KaHip is interesting.
FlowCutter gets slower with a decreasing maximum imbalance.
However, KaHip gets slower with an increasing maximum imbalance, i.e., the other way round.
A clear ranking is therefore not possible but the tendency for max imbalances above 10\% is that F3 is the fastest, followed by K0.73, followed by F20, and finally K1.00.

\subsubsection{Discussion for Central Europe}

\makeparetotable{pareto/central_eur2}{\label{tab:pareto-ceur}Results for the DIMACS Central Europe graph.}

\paragraph{Cut sizes}
In Table~\ref{tab:pareto-ceur} we report the results of our experiments for the Central Europe graph.
The most striking observation is that the cut sizes in this graph are larger than those in any of the USA graphs.
This explains why the Europe graph has a higher tree-width and larger search spaces than the USA graph.
It is not immediately clear which cut is the best for our application, however, the cuts with sizes 180, 162, and 155 seem to offer a good trade-off between cut size and imbalance.
F20 manages to find all of them. 
K1.00 finds a variant of the 162 edge cut with a marginally higher imbalance.
InertialFlow is able of finding the 162 and the 155 edge cuts.
Unfortunately, as already previously observed we need to set the $\max \epsilon$ parameter significantly higher than the imbalance of the cuts for InertialFlow to find them.

\paragraph{Running Times}
On all of the USA graphs F20 was at least on par with K1.00 in terms of running time and often even faster.
On this graph we see a significant gap of at least a factor 2 for all imbalances below $70\%$.
The explanation is that the running time of FlowCutter does not only depend on the graph size but also on the cut size.
As this graph has larger cuts than the USA graphs, FlowCutter is slower.
KaHip's running time is not or at least less affected by cut size and therefore comes out ahead on this graph.
However, for our particular application, i.e., nested dissection, a running time sensitive to the cut size is a good thing.
We have only few top level cuts with large cuts but many more low level cuts that have tiny cuts.
A partitioner that gets faster, the smaller the cuts become is therefore useful in this scenario as it gets faster on the lower levels.
This observation also explains why F20 wins against K1.00 in terms of running time on the Europe graph in the CCH experiment.

\subsection{Special Structure of the Europe Graph}
\label{sec:europe-graph}

\makeparetotable{pareto/europe}{\label{tab:pareto-eur}Results for the DIMACS Europe.}

\begin{figure}
\begin{center}
\subfloat[\label{fig:eur-kahip}K0.76]{\begin{centering}
\includegraphics[width=3.5cm]{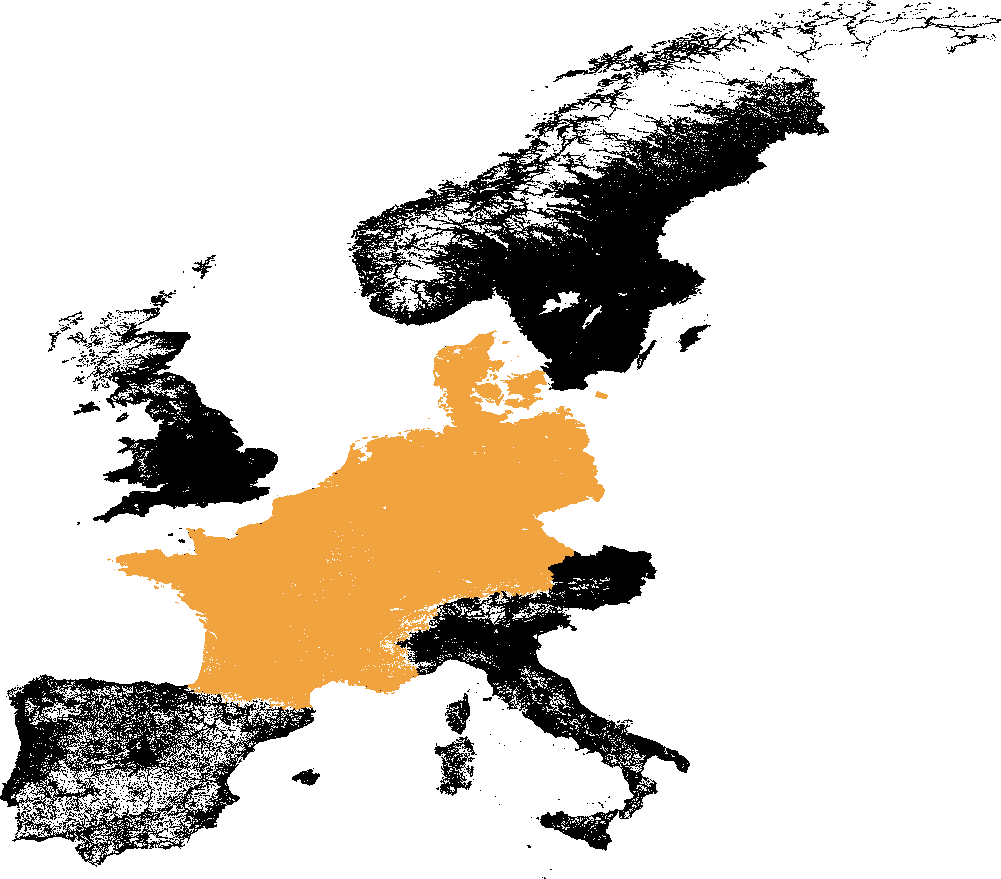}
\end{centering}

} \subfloat[\label{fig:eur-flowcutter-sat}F with guidance]{\begin{centering}
\includegraphics[width=3.5cm]{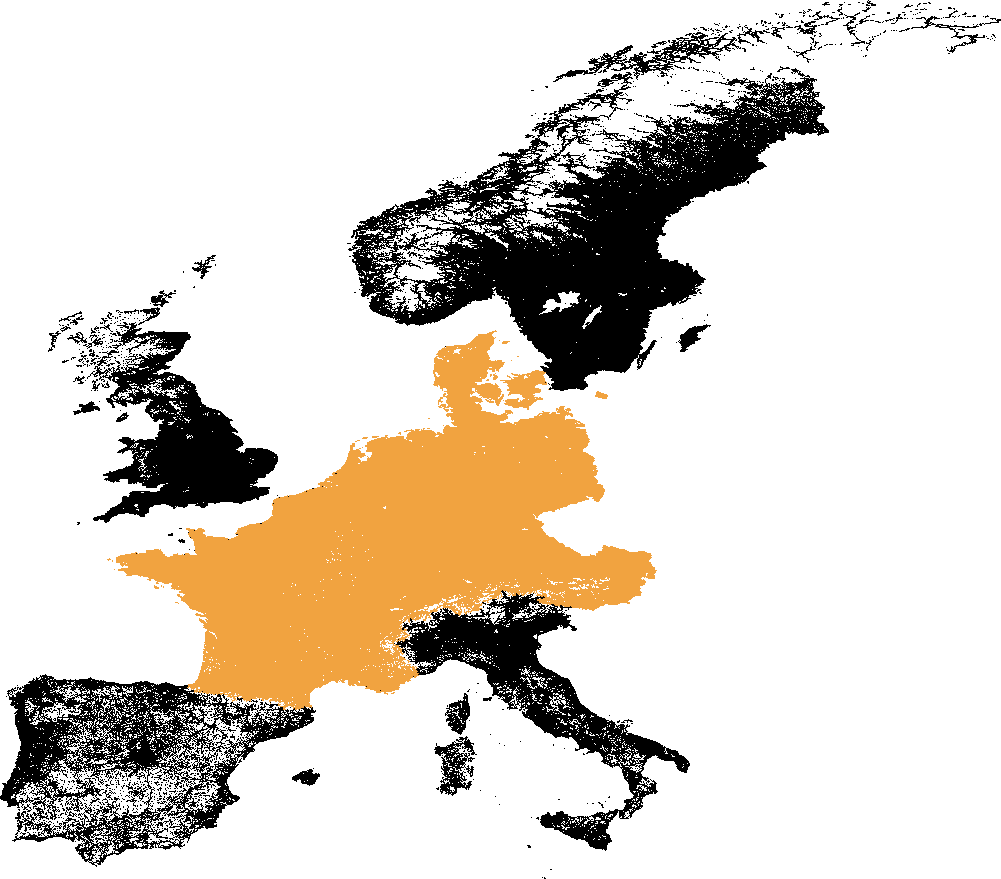}
\end{centering}

} \subfloat[\label{fig:eur_flowcutter_rhine}F20]{\begin{centering}
\includegraphics[width=3.5cm]{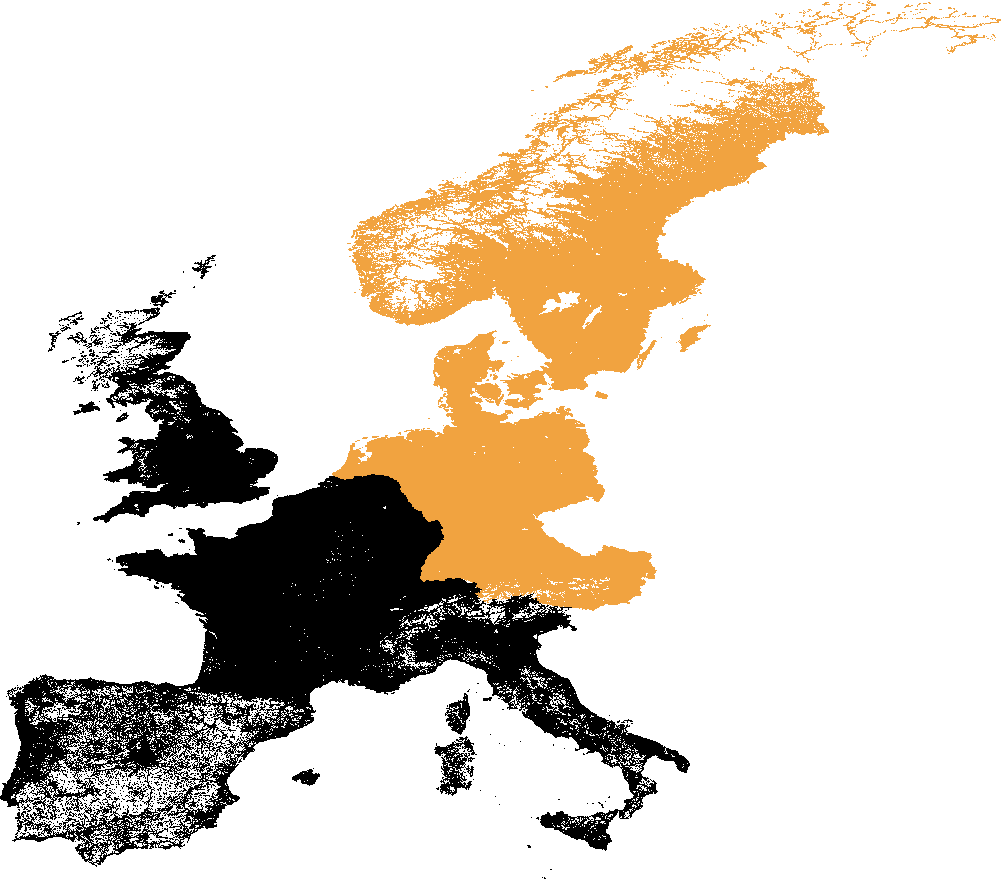}
\end{centering}

}
\end{center}

\caption{\label{fig:eur-cuts}Various top-level Europe cuts.}

\end{figure}

We present the results for the Europe graph in Table~\ref{tab:pareto-eur}.
The reported cut sizes do not follow the pattern observed on the other graphs.
The cuts of F20 are significantly larger than those of KaHip.
Another observation is that most of the cuts found by partitioners except FlowCutter do not have connected sides.
This already hints at the root of the problem.
In Figure~\ref{fig:eur-cuts} we visualized the cuts found. 
Figure~\ref{fig:eur-kahip} depicts the cut found by KaHip with 112 edges and Figure~\ref{fig:eur_flowcutter_rhine} depicts the cut found by F20 with 188 edges.
Visually these two cuts look very different. 
To explain the effect in detail we must first describe some properties of the Europe graph.

\paragraph{Unique Geography}

Top-level Europe has a unique geographic topology. 
There is a well connected center formed by France, Germany, Belgium, Luxembourg, and the Netherlands.
Further, there are four peninsulas.
Spain and Portugal are only connected by a comparatively small piece of land with France.
Italy is separated by the rest of Europe by the Alps.
Sweden and Norway are separated by the Baltic Sea from Central Europe. 
They are only connected to Denmark by a highway bridge in Kopenhagen. 
This bridge is also the cut with 2 edges with 72\% imbalance found by several partitioners.
Great Britain is separated by the North Sea and is only connected to the continent using ferries, which are treated as roads in the benchmark dataset.
There are further ferries between Spain and England, and between Spain and Italy.
However, there are no ferries from or to Scandinavia.

\paragraph{Structure of KaHip Cuts}

The KaHip cut with 112 edges separates Central Europe from its peninsulas. 
The sides are not connected because there is no path from Great Britain to Scandinavia.
The KaHip cut with 129 edges with connected sides further separates Denmark from Germany.
The sides of the cut are connected, as there is a ferry from England to Denmark and a bridge from Denmark to Sweden.

\paragraph{Structure of FlowCutter Cuts}

The FlowCutter cut is structurally very different.
FlowCutter separates Central Europe along the Rhine river and the Alps.
FlowCutter cannot find the 112 edge cut because its sides are not connected.
Further, it does not find the 129 edge cut because the shape of this cut is very different from what the employed piercing heuristic expects.

\paragraph{KaHip vs FlowCutter}

At first glance, KaHip seems to be better than FlowCutter on this instance.
However, this is not consistent with our observation that FlowCutter produces better contraction orders.
The explanation is that, as we consider a recursive bisection, the question is not whether Central Europe must be cut along the Rhine river, but at which recursion level we do it.
FlowCutter does it at the top level, whereas KaHip does it at a lower level.
It is unclear which approach is better. 
We will investigate this question in detail below.
However, before we answer this question we explore how we can modify FlowCutter to find a cut similar to the one found by KaHip.

\paragraph{Adapting FlowCutter}

One can regard the balanced 112 edge cut of KaHip as union of four smaller edge cuts with a higher imbalances.
There is one cut for each peninsula.
Repeatedly cutting of each peninsula on consecutive levels of the recursion is equivalent with cutting them all in one level.
The question is therefore whether FlowCutter is able to find one of the peninsula cuts and this is indeed the case.
FlowCutter finds the cut with 2 edges that separates Scandinavia from the rest. 
However, FlowCutter refrains from choosing this cut from the Pareto-set because we have a hard bound on a maximum imbalance of at most 60\%.

\begin{table}
\begin{center}
\begin{tabular}{crrc}
\toprule 
 & Lat & Lon & Place\\
\midrule 
Source & 49.0 & 8.4 & Karlsruhe\\
\midrule 
\multirow{4}{*}{Target} & 41.0 & 16.9 & Bari\\
 & 38.7 & -9.1 & Lisbon\\
 & 53.5 & -2.8 & Liverpool\\
 & 59.2 & 18.0 & Stockholm\\
\bottomrule 
\end{tabular}
\end{center}

\caption{\label{tab:handpicked-st}Handpicked source and target nodes.}
\end{table}%

Another option to help FlowCutter is to handpick source and target nodes.
We selected the nodes which are closed to the coordinates given in Table~\ref{tab:handpicked-st} and used these as input to FlowCutter.
These coordinates are not magic numbers. 
They represent positions chosen at the extremities of the peninsulas and in the center of Central Europe.
Most humans are able to deduce this information from looking at a Europe map.
With this setup we were not able to find the 112 edge cut with 7.9\% imbalance found by KaHip. 
However, we were able to find another cut with a seemingly better trade-off.
This new cut is depicted in~\ref{fig:eur-flowcutter-sat}.
It has 87 edges and 15\% imbalance.
The smaller cut results from placing Austria on the other side of the cut compared to the 112 edge cut of KaHip and from some minor improvements along the other borders.
KaHip is incapable of finding this cut.

\begin{table}

\setlength\tabcolsep{3.5pt}

\begin{center}
\begin{tabular}{crrrrrrrrrr}
\toprule
  & \multicolumn{4}{c}{Search Space} & \#Arcs  &  & Up. \\
\cmidrule(lr){2-5} 
  & \multicolumn{2}{c}{Nodes} & \multicolumn{2}{c}{Arcs {[}$\cdot10^{3}${]}} & in CCH & \#Tri. & Tw.\\
\cmidrule(lr){2-3}  \cmidrule(lr){4-5} 
 & Avg. & Max. & Avg. & Max. & {[}$\cdot10^{6}${]} & {[}$\cdot10^{6}${]} & Bd. \\
\midrule
F3 & 734.1 & 1\,159 & 140.2 & 312 & 60.3 & 519.4 & 531 \\
F20 & 616.0 & 1\,102 & 102.8 & 268 & \textbf{58.8} & 459.6 & 455 \\
F100 & 622.6 & 1\,105 & 104.8 & \textbf{239} & \textbf{58.8} & 459.4 & 449 \\
\midrule
F3+H & 625.1 & 1\,151 & 106.2 & 262 & 60.2 & 509.2 & \textbf{439} \\
F20+H & 601.2 & \textbf{1\,064} & 98.9 & 261 & \textbf{58.8} & 456.9 & 444 \\
F100+H & \textbf{600.6} & 1\,065 & \textbf{98.6} & 250 & \textbf{58.8} & \textbf{454.4} & 444 \\
\bottomrule
\end{tabular}
\end{center}

\caption{\label{tab:Contraction-Orders-Europe}Contraction Order Experiments on the DIMACS Europe graph.
F3, F20, F100 are the default FlowCutter variants that use a top-level cut along the Rhine river.
F3+H, F20+H, F100+H use a handpicked top-level cut separating Central Europe from the peninsulas.
}

\end{table}

\paragraph{The Best Top-level Cut}

We have shown that with a bit of help it is possible to push FlowCutter towards computing a small cut that separates the peninsulas.
Now, we will answer the question whether this a better top-level cut than the 188 edge cut found by the default FlowCutter configuration.
We derive an 87 node separator from the 87 edge cut and place these nodes at the end of the contraction order manually.
We then run FlowCutter on the resulting sides recursively without any further manual guidance.
In Table~\ref{tab:Contraction-Orders-Europe} we report the characteristics of the so obtained orders.
The new orders are marked with ``+H'', indicating human interaction. 
We compare them with the default FlowCutter orders. 
The new orders seem to be slightly superior with respect to every criteria except the maximum number of arcs in the search space where the default FlowCutter orders seem to win.
Further, the orders seem to produce a similar number of edges in the CCH regardless of the top-level cut used.
However, the differences in order quality are very minor.
We observe with respect to no criterion a difference that is larger than 2\%.
This difference can be due to a peninsula top-level cut being slightly better.
However, another explanation is that FlowCutter finds better cuts on the lower levels because a difficult to find peninsula cut was eliminated manually.
In either case, the differences are so small that we decided that it is not worthwhile to automatize the selection of a top level peninsula cut.

\subsection{Walshaw Benchmark Set}
\label{app:walshaw}

\begin{table}
\setlength\tabcolsep{3.5pt}

\begin{center}
\begin{tabular}{clrrrrr}
\toprule 
 &  & \multicolumn{4}{c}{minimum edges in cut for } & running\\
\cmidrule(lr){3-6}
graph & algorithm & $\epsilon=0\%$ & $\epsilon=1\%$ & $\epsilon=3\%$ & $\epsilon=5\%$ & time {[}s{]}\\
\midrule
144 & F20 & 6\,649 & 6\,608 & 6\,514 & 6\,472 & 2\,423.82\\
144K nodes & F100 & 6\,515 & 6\,479 & 6\,456 & 6\,366 & 10\,437.91\\
1074K edges & Reference & \textbf{6\,486} & \textbf{6\,478} & \textbf{6\,432} & \textbf{6\,345} & \\
\midrule
3elt & F20 & \textbf{90*} & \textbf{89} & \textbf{87} & \textbf{87} & 0.36\\
4720 nodes & F100 & \textbf{90*} & \textbf{89} & \textbf{87} & \textbf{87} & 1.87\\
13K edges & Reference & \textbf{90*} & \textbf{89} & \textbf{87} & \textbf{87} & \\
\midrule
4elt & F20 & 149 & \textbf{138} & \textbf{137} & \textbf{137} & 1.97\\
15K nodes & F100 & \textbf{139*} & \textbf{138} & \textbf{137} & \textbf{137} & 9.50\\
45K edges & Reference & \textbf{139*} & \textbf{138} & \textbf{137} & \textbf{137} & \\
\midrule
598a & F20 & 2\,417 & 2\,390 & \textbf{2\,367} & \textbf{2\,336} & 545.69\\
110K nodes & F100 & 2\,400 & \textbf{2\,388} & \textbf{2\,367} & \textbf{2\,336} & 2\,675.32\\
741K edges & Reference & \textbf{2\,398} & \textbf{2\,388} & \textbf{2\,367} & \textbf{2\,336} & \\
\midrule
auto & F20 & 10\,609 & 10\,283 & 9\,890 & \textbf{9\,450} & 13\,445.66\\
448K nodes & F100 & 10\,549 & 10\,283 & 9\,823 & \textbf{9\,450} & 66\,249.82\\
3314K edges & Reference & \textbf{10\,103} & \textbf{9\,949} & \textbf{9\,673} & \textbf{9\,450} & \\
\midrule
bcsstk30 & F20 & 6\,454 & 6\,347 & \textbf{6\,251} & \textbf{6\,251} & 245.65\\
28K nodes & F100 & 6\,408 & 6\,347 & \textbf{6\,251} & \textbf{6\,251} & 1\,230.27\\
1007K edges & Reference & \textbf{6\,394} & \textbf{6\,335} & \textbf{6\,251} & \textbf{6\,251} & \\
\midrule
bcsstk33 & F20 & 10\,220 & \textbf{10\,097} & \textbf{10\,064} & \textbf{9\,914} & 118.38\\
8738 nodes & F100 & 10\,177 & \textbf{10\,097} & \textbf{10\,064} & \textbf{9\,914} & 573.02\\
291K edges & Reference & \textbf{10\,171} & \textbf{10\,097} & \textbf{10\,064} & \textbf{9\,914} & \\
\midrule
brack2 & F20 & 742 & \textbf{708} & \textbf{684} & \textbf{660} & 58.13\\
62K nodes & F100 & 742 & \textbf{708} & \textbf{684} & \textbf{660} & 283.99\\
366K edges & Reference & \textbf{731*} & \textbf{708} & \textbf{684} & \textbf{660} & \\
\midrule
crack & F20 & \textbf{184} & \textbf{183} & \textbf{182} & \textbf{182} & 2.17\\
10K nodes & F100 & \textbf{184} & \textbf{183} & \textbf{182} & \textbf{182} & 10.97\\
30K edges & Reference & \textbf{184} & \textbf{183} & \textbf{182} & \textbf{182} & \\
\midrule
cs4 & F20 & 381 & 371 & 367 & 360 & 11.68\\
22K nodes & F100 & 372 & 370 & 365 & 357 & 58.11\\
43K edges & Reference & \textbf{369} & \textbf{366} & \textbf{360} & \textbf{353} & \\
\midrule
cti & F20 & 342 & \textbf{318} & \textbf{318} & \textbf{318} & 6.10\\
16K nodes & F100 & 339 & \textbf{318} & \textbf{318} & \textbf{318} & 30.55\\
48K edges & Reference & 334 & \textbf{318} & \textbf{318} & \textbf{318} & \\
\midrule
fe\_4elt2 & F20 & \textbf{130*} & \textbf{130} & \textbf{130} & \textbf{130} & 1.86\\
11K nodes & F100 & \textbf{130*} & \textbf{130} & \textbf{130} & \textbf{130} & 9.19\\
32K edges & Reference & \textbf{130*} & \textbf{130} & \textbf{130} & \textbf{130} & \\
\bottomrule
\end{tabular}
\end{center}

\caption{\label{tab:Walshaw-part1}
Performance on the Walshaw benchmark set, part 1. 
``Reference'' is the best known bisection for the graph as maintained by Walshaw.
A ``*'' marks solutions for which optimality has been shown.
}
\end{table}

\begin{table}

\setlength\tabcolsep{3.5pt}

\begin{center}
\begin{tabular}{clrrrrr}
\toprule 
 &  & \multicolumn{4}{c}{minimum edges in cut for } & running\\
\cmidrule(lr){3-6}
graph & algorithm & $\epsilon=0\%$ & $\epsilon=1\%$ & $\epsilon=3\%$ & $\epsilon=5\%$ & time {[}s{]}\\
\midrule
fe\_ocean & FlowCutter 20 & 504 & 431 & \textbf{311} & \textbf{311} & 89.70\\
143K nodes & FlowCutter 100 & 483 & 408 & \textbf{311} & \textbf{311} & 418.60\\
409K edges & Reference & \textbf{464} & \textbf{387} & \textbf{311} & \textbf{311} & \\
\midrule
fe\_rotor & FlowCutter 20 & 2\,115 & 2\,091 & \textbf{1\,959} & 1\,948 & 334.58\\
99K nodes & FlowCutter 100 & 2\,106 & 2\,067 & \textbf{1\,959} & \textbf{1\,940} & 1\,636.78\\
662K edges & Reference & \textbf{2\,098} & \textbf{2\,031} & \textbf{1\,959} & \textbf{1\,940} & \\
\midrule
fe\_sphere & FlowCutter 20 & \textbf{386} & \textbf{386} & \textbf{384} & \textbf{384} & 5.98\\
16K nodes & FlowCutter 100 & \textbf{386} & \textbf{386} & \textbf{384} & \textbf{384} & 30.84\\
49K edges & Reference & \textbf{386} & \textbf{386} & \textbf{384} & \textbf{384} & \\
\midrule
fe\_tooth & FlowCutter 20 & 3\,852 & 3\,841 & 3\,814 & \textbf{3\,773} & 413.48\\
78K nodes & FlowCutter 100 & 3\,836 & 3\,832 & 3\,790 & \textbf{3\,773} & 2\,067.54\\
452K edges & Reference & \textbf{3\,816} & \textbf{3\,814} & \textbf{3\,788} & \textbf{3\,773} & \\
\midrule
finan512 & FlowCutter 20 & \textbf{162*} & \textbf{162} & \textbf{162} & \textbf{162} & 8.11\\
74K nodes & FlowCutter 100 & \textbf{162*} & \textbf{162} & \textbf{162} & \textbf{162} & 39.01\\
261K edges & Reference & \textbf{162*} & \textbf{162} & \textbf{162} & \textbf{162} & \\
\midrule
m14b & FlowCutter 20 & 3\,858 & \textbf{3\,826} & \textbf{3\,823} & 3\,805 & 2\,115.07\\
214K nodes & FlowCutter 100 & \textbf{3\,836} & \textbf{3\,826} & \textbf{3\,823} & 3\,804 & 10\,512.24\\
1679K edges & Reference & \textbf{3\,836} & \textbf{3\,826} & \textbf{3\,823} & \textbf{3\,802} & \\
\midrule
t60k & FlowCutter 20 & 80 & 79 & 73 & \textbf{65} & 2.98\\
60K nodes & FlowCutter 100 & 80 & 77 & \textbf{71} & \textbf{65} & 14.55\\
89K edges & Reference & \textbf{79} & \textbf{75} & \textbf{71} & \textbf{65} & \\
\midrule
vibrobox & FlowCutter 20 & 10\,614 & 10\,356 & 10\,356 & 10\,356 & 139.90\\
12K nodes & FlowCutter 100 & 10\,365 & \textbf{10\,310} & \textbf{10\,310} & \textbf{10\,310} & 680.76\\
165K edges & Reference & \textbf{10\,343} & \textbf{10\,310} & \textbf{10\,310} & \textbf{10\,310} & \\
\midrule
wave & FlowCutter 20 & 8\,734 & 8\,734 & 8\,734 & 8\,724 & 2\,723.12\\
156K nodes & FlowCutter 100 & 8\,716 & 8\,673 & 8\,650 & 8\,590 & 13\,583.59\\
1059K edges & Reference & \textbf{8\,677} & \textbf{8\,657} & \textbf{8\,591} & \textbf{8\,524} & \\
\midrule
whitaker3 & FlowCutter 20 & \textbf{127*} & \textbf{126} & \textbf{126} & \textbf{126} & 1.49\\
9800 nodes & FlowCutter 100 & \textbf{127*} & \textbf{126} & \textbf{126} & \textbf{126} & 7.00\\
28K edges & Reference & \textbf{127*} & \textbf{126} & \textbf{126} & \textbf{126} & \\
\midrule
wing & FlowCutter 20 & 790 & 790 & 790 & 790 & 80.11\\
62K nodes & FlowCutter 100 & 790 & 790 & 781 & 773 & 401.82\\
121K edges & Reference & \textbf{789} & \textbf{784} & \textbf{773} & \textbf{770} & \\
\midrule
wing\_nodal & FlowCutter 20 & 1\,767 & 1\,764 & 1\,715 & 1\,691 & 27.02\\
10K nodes & FlowCutter 100 & 1\,743 & 1\,740 & 1\,710 & 1\,688 & 134.05\\
75K edges & Reference & \textbf{1\,707} & \textbf{1\,695} & \textbf{1\,678} & \textbf{1\,668} & \\
\bottomrule
\end{tabular}
\end{center}

\caption{\label{tab:Walshaw-part2}Performance on the Walshaw benchmark set, part 2.}

\end{table}

A popular set of graph partitioning benchmark instances is maintained by Walshaw~\cite{swc-acesm-04}. 
The data contains 34 graphs and solutions to the edge-bisection problem with non-connected sides and maximum imbalance values of $\epsilon=0\%$, $\epsilon=1\%$, $\epsilon=3\%$, and $\epsilon=5\%$. 
These archived solutions are the best cuts that any partitioner has found so far. 
A few of them were even proven to be optimal \cite{dfgrw-a-14}. 
Comparing against these archived solutions allows us to compare FlowCutter quality-wise against the state of the art.
We want to stress that this state of the art was computed by a large mixture of algorithms with an even larger set of parameters that may have been chosen in instance-dependent ways. 
We compare this against a single algorithm with a single set of parameters. 
Further FlowCutter was designed for higher imbalances than 5\%. 
It was not tuned for the cases with a lower imbalance. 
FlowCutter only computes cuts with connected sides. 
We therefore filter out all graphs that are either not connected or where the archived $\epsilon=0$-solution has non-connected sides. 
Of the 34 graphs only 24 remain. 
The results are reported in Tables~\ref{tab:Walshaw-part1} and~\ref{tab:Walshaw-part2}.

For $\epsilon=5\%$ there are only six graphs where FlowCutter does not match the best known cut quality. 
These are: ``144'', ``cs4'', ``m14b'', ``wave'', ``wing'', and ``wing\_nodal''. 
For three of these graphs, FlowCutter finds cuts that are larger by a negligible amount of at most 5 edges. 
For the other three, the cuts found are larger but are still close to the best known solutions. 
For lower imbalances, the results are not quite as good but still very close to the best known solutions. 

In terms of running time the results are more mixed. 
Some cuts are found very quickly, while FlowCutter needs a significant amount of time on others. 
This is due to the fact that its running time is in $O(cm)$. 
If both, the cut size $c$ and the edge count $m$, are large, then $O(cm)$ is large. 
However, for graphs with small cuts the algorithm scales nearly linearly in the graph size.

\section{Conclusion}

We introduce FlowCutter, a graph bisection algorithm that optimizes balance and cut size in the Pareto sense. 
The core algorithm computes small, balanced edge cuts separating two input nodes $s$ and $t$.
Upon this core algorithm, we build algorithms to compute overall small, balanced edges cuts independent of an $st$-pair specified in the input.
We further extend our algorithm to compute small balanced node separators.
By combining FlowCutter with a nested dissection-based strategy, we compute contraction orders (also called elimination or minimum fill-in orders). 
We show that our orders beat the state-of-the-art in terms of quality on road graphs.
We evaluate the quality of our orders by directly applying them in the context of Customizable Contraction Hierarchies, a speedup technique for shortest paths. 
Further, we show that FlowCutter manages to equate the best known cuts for many instances of the Walshaw benchmark set, demonstrating that FlowCutter is applicable beyond just bisecting road graphs.
Finally, we use FlowCutter to compute tree-decompositions of small width.
To evaluate the performance of our method, we submitted FlowCutter to the PACE2016 challenge~\cite{dhjkkr-tfpac-16}, where it won the first place in the corresponding sequential track.
This demonstrates that FlowCutter works well on a broad class of graphs.
The source code of the PACE 2016 submission is available at~\cite{pace-code}. 

\paragraph{Future Work}

We show that FlowCutter is an excellent tool to be used within shortest path acceleration techniques on road graphs.
Luckily, the developed techniques seem also useful in other domains. 
Experimental evaluations that take domain specific requirements into account would be interesting venue for future research. 
For example, our secondary piercing heuristic could be swapped out with one that uses information that is only available in certain applications. 

Another open question is how to adapt FlowCutter to graphs that have weighted nodes or weighted edges or even both.

We use a nested dissection scheme to compute contraction orders with FlowCutter as bisection algorithm.
This seems somewhat wasteful as in this setup FlowCutter computes a large set of cuts but we only use one of them and then discard the others. 
In the deeper levels of the recursion FlowCutter will then likely recompute some of the discarded cuts.
Adapting the nested dissection scheme in a way that utilizes several cuts from each set could significantly improve the running time. 

FlowCutter needs two initial nodes on separate sides of the cut. 
Currently these are determined by random sampling. 
A better selection strategy could decrease the number of samples needed. 

\textbf{Acknowledgment:} We thank Roland Glantz for helpful discussions.


\end{document}